\documentclass[final]{llncs}

\usepackage{color}
\usepackage{graphicx}
\usepackage[pdfusetitle]{hyperref}
\usepackage{cite}
\usepackage{amsmath,amssymb}
\usepackage{wrapfig}
\usepackage[calc]{adjustbox}
\newlength{\strutheight}
\usepackage{stmaryrd}
\usepackage{mathtools}
\mathtoolsset{showonlyrefs=true}
\usepackage{tikz,pgfplots}
\usetikzlibrary{cd}
\tikzset{%
    symbol/.style={%
        draw=none,
        every to/.append style={%
            edge node={node [sloped, allow upside down, auto=false]{$#1$}}}
    }
}
\usepackage{bussproofs}
\usepackage{mathpartir}

\usepackage{ifthen}
\newboolean{longversion}
\ifthenelse{\equal{\detokenize{long}}{\jobname}}{
	\setboolean{longversion}{true}
}{
	\setboolean{longversion}{false}
}
\newcommand{\referappendix}[2]{\ifthenelse{\boolean{longversion}}{\S\ref{#1}}{\cite[Appendix {#2}]{}}}

\spnewtheorem{mytheorem}{Theorem}{\bfseries}{\itshape}
\spnewtheorem{mylemma}[mytheorem]{Lemma}{\bfseries}{\itshape}
\spnewtheorem{myproposition}[mytheorem]{Proposition}{\bfseries}{\itshape}
\spnewtheorem{mysublemma}[mytheorem]{Sublemma}{\bfseries}{\itshape}
\spnewtheorem{mycorollary}[mytheorem]{Corollary}{\bfseries}{\itshape}
\spnewtheorem{myfact}[mytheorem]{Fact}{\bfseries}{\itshape}
\spnewtheorem{mynotation}[mytheorem]{Notation}{\bfseries}{\rmfamily}
\spnewtheorem{myremark}[mytheorem]{Remark}{\bfseries}{\rmfamily}
\spnewtheorem{myexample}[mytheorem]{Example}{\bfseries}{\rmfamily}
\spnewtheorem{myassumption}[mytheorem]{Assumption}{\bfseries}{\rmfamily}
\spnewtheorem{mydefinition}[mytheorem]{Definition}{\bfseries}{\rmfamily}
\spnewtheorem{myrequirements}[mytheorem]{Requirements}{\bfseries}{\rmfamily}
\spnewtheorem{myproblem}[mytheorem]{Problem}{\bfseries}{\rmfamily}
\spnewtheorem{myconjecture}[mytheorem]{Conjecture}{\bfseries}{\rmfamily}

\usepackage{macros.basic}
\newcommand{\identity}[1]{\mathrm{id}_{#1}}

\newcommand{\pullback}[1]{#1^{*}}
\newcommand{\CartesianOver}[2]{\overline{#1}({#2})}
\newcommand{\coreindexing}[1]{#1_{!}}
\newcommand{\CoCartesianOver}[2]{\underline{#1}(#2)}

\newcommand{\ComprehensionFunctor}[1]{\{ #1 \}}

\newcommand{\CPO}{\mathbf{CPO}}
\newcommand{\famset}{\mathsf{fam}_{\Set}}
\newcommand{\simple}[1]{\mathsf{s}_{#1}}
\newcommand{\subobj}[1]{\mathsf{sub}_{#1}}

\newcommand{\RefinedCCompC}[2]{{\{ #1 \mid #2 \}}}
\newcommand{\refinedccompc}[2]{{\{ #1 \mid #2 \}}}

\newcommand{\underlying}[1]{#1_{u}}
\newcommand{\emptyctx}{\diamond}
\newcommand{\ElimRefinement}[1]{|#1|}
\newcommand{\subtype}{<:}
\newcommand{\WellFormedContext}[1]{\vdash #1}
\newcommand{\WellFormedType}[2]{#1 \vdash #2}
\newcommand{\WellTypedTerm}[3]{#1 \vdash #2 : #3}
\newcommand{\Subtyping}[3]{#1 \vdash #2 \subtype #3}
\newcommand{\TermEqual}[4]{#1 \vdash #2 = #3 : #4}
\newcommand{\ContextSubtyping}[2]{\vdash #1 \subtype #2}
\newcommand{\ContextEqual}[2]{\vdash #1 = #2}
\newcommand{\TypeEqual}[3]{#1 \vdash #2 = #3}

\newcommand{\ValueSumType}[3]{\Sigma #1 {:} #2. #3}
\newcommand{\CompProdType}[3]{\Pi #1 {:} #2. #3}

\newcommand{\thunk}[1]{\mathbf{thunk}\ #1}
\newcommand{\force}[2]{\mathbf{force}_{#1}\ #2}
\newcommand{\return}[1]{\mathbf{return}\ #1}
\newcommand{\SeqComp}[5]{#1\ \mathbf{to}\ #2 : #3\ \mathbf{in}_{#4}\ #5}
\newcommand{\LambdaAbs}[3]{\lambda #1 {:} #2. #3}
\newcommand{\App}[5]{#1 (#2)_{(#3 {:} #4). #5}}
\newcommand{\DTuple}[5]{\langle #1, #2 \rangle_{(#3 : #4). #5}}
\newcommand{\PatternMatch}[8]{\mathbf{pm}\ #1\ \mathbf{as}\ \langle #2 : #3, #4 : #5 \rangle\ \mathbf{in}_{#6. #7}\ #8}
\newcommand{\Case}[9]{\mathbf{case}\ #1\ \mathbf{of}_{#2.#3}\ (\mathbf{inl}\ (#4 : #5) \mapsto #6, \mathbf{inr}\ (#7 : #8) \mapsto #9)}

\title{General Semantic Construction of Dependent Refinement Type Systems, Categorically}
\author{Satoshi Kura\inst{1,2}\orcidID{0000-0002-3954-8255}}
\institute{National Institute of Informatics, Tokyo, Japan
\and The Graduate University for Advanced Studies (SOKENDAI), Kanagawa, Japan}

\begin{document}
\maketitle

\begin{abstract}
    Refinement types are types equipped with predicates that specify preconditions and postconditions of underlying functional languages.
    We propose a general semantic construction of dependent refinement type systems from underlying type systems and predicate logic, that is, a construction of liftings of closed comprehension categories from given (underlying) closed comprehension categories and posetal fibrations for predicate logic.
    We give sufficient conditions to lift structures such as dependent products, dependent sums, computational effects, and recursion from the underlying type systems to refinement type systems.
    We demonstrate the usage of our construction by giving semantics to a refinement type system and proving soundness.
\end{abstract}

\section{Introduction}
% What is refinement types
Refinement types~\cite{freeman1991,flanagan2006} are types equipped with predicates that restrict values in the types.
They are used to specify preconditions and postconditions and to verify that programs satisfy the specifications.
Refinement type systems are dependently typed to express postconditions of functions that depend on input.
% Examples
Many dependent refinement types systems are proposed~\cite{flanagan2006,knowles2008,barthe2015,lehmann2017,unno2018} and implemented in, e.g., F${}^\star$~\cite{swamy2013,swamy2013a} and LiquidHaskell~\cite{rondon2008,vazou2013,vazou2014}.

% relation between dependent refinement type system, underlying type system, and predicate logic
%Refinement type systems are combinations of underlying type systems and predicate logic.
%Refinement type systems inherit terms of underlying type systems, while types are refined by predicate logic and have subtyping relations induced by logical implication.
%Refinement type systems are dependently typed even if underlying type systems are simply typed.

In this paper, we address the question: ``How are refinement type systems, underlying type systems, and predicate logic related from the viewpoint of categorical semantics?''
Although most existing refinement type systems are proved to be sound using operational semantics,
we believe that categorical semantics is more suitable for the general understanding of their nature, especially when we consider general computational effects and various kinds of predicate logic (e.g., for relational verification).
This understanding will provide guidelines to design new refinement type systems.

Our answer to the question is a general semantic construction of refinement type systems from underlying type systems and predicate logic.
More concretely, given a closed comprehension category (CCompC for short) for interpreting an underlying type system and a fibration for predicate logic, we combine them to obtain another CCompC that can interpret a refinement type system built from the underlying type system and the predicate logic.

For example, consider giving an interpretation to the term ``$x : \{ \mathrm{int} \mid x \ge 0 \} \vdash x + 1 : \{ v : \mathrm{int} \mid v = x + 1 \}$'' in a refinement type system.
Its underlying term is ``$x : \mathrm{int} \vdash x + 1 : \mathrm{int}$,'' and we assume that it is interpreted as the successor function of $\mathbb{Z}$ in $\Set$.
The problem here is how to refine this interpretation with predicates.
In dependent refinement types, predicates may depend on the variables in contexts.
In this example, the type ``$x : \{ \mathrm{int} \mid x \ge 0 \} \vdash \{ v : \mathrm{int} \mid v = x + 1 \}$'' depends on the variable $x$.
Thus, the interpretation of such types must be a predicate on the context and the type, i.e., 
\[ \llbracket x : \{ \mathrm{int} \mid x \ge 0 \} \vdash \{ v : \mathrm{int} \mid v = x + 1 \} \rrbracket = \{ (x, v) \in \mathbb{Z} \times \mathbb{Z} \mid x \ge 0 \land v = x + 1 \}. \]
As a result, the term in the refinement type system is interpreted as the interpretation in the underlying type system together with the property that if input satisfies preconditions, then output satisfies postconditions.
\begin{equation}
	\begin{tikzcd}[row sep=small]
		\{ x \in \mathbb{Z} \mid x \ge 0 \} \ar[r, dashed] \ar[d, symbol=\subseteq] & \{ (x, v) \in \mathbb{Z} \times \mathbb{Z} \mid x \ge 0 \land v = x + 1 \} \ar[d, symbol=\subseteq] \\
		\mathbb{Z} \ar[r, "{\langle \identity{\mathbb{Z}}, ({-}) + 1 \rangle}"] & \mathbb{Z} \times \mathbb{Z}
	\end{tikzcd} \label{eq:example_denotation}
\end{equation}

\begin{wrapfigure}{r}{7em}
	\vspace*{-2.5em}
	\begin{tikzcd}[row sep=small, column sep=small]
		\RefinedCCompC{\category{E}}{\category{P}} \ar[r] \ar[d] & \category{E} \ar[d, "p"] \\
		\category{P} \ar[r, "q"] & \category{B}
	\end{tikzcd}
	\caption{Lifting.}
	\label{fig:lifting_intro}
	\vspace*{-2em}
\end{wrapfigure}
We formalize this refinement process as a construction of liftings of CCompCs, which are used to interpret dependent type theories.
Assume that we have a pair of a CCompC $p : \category{E} \to \category{B}$ for interpreting underlying type systems and a fibration $q : \category{P} \to \category{B}$ for predicate logic satisfying certain conditions.
Then we construct a CCompC $\RefinedCCompC{\category{E}}{\category{P}} \to \category{P}$ for interpreting dependent refinement type systems.
This construction also yields a morphism of CCompCs from $\RefinedCCompC{\category{E}}{\category{P}} \to \category{P}$ to $p : \category{E} \to \category{B}$ in Fig.~\ref{fig:lifting_intro}.
Given the simple fibration $\mathbf{s}(\Set) \to \Set$ for underlying type systems and the subobject fibration $\mathbf{Sub}(\Set) \to \Set$ for predicate logic, then we get interpretations like~\eqref{eq:example_denotation}.
%Objects in $\{ p \mid q \}$ are defined by the tuple $((I, X), P, Q)$ where $I$ represents a context, $X$ a type, $P$ a predicate on $I$, and $Q$ a predicate on $I \times X$; and the functor $\{ p \mid q \} \to \category{P}$ is defined by $((I, X), P, Q) \mapsto P$.

We extend the construction of liftings of CCompCs to liftings of fibred monads on CCompCs, which is motivated by the fact that many dependent refinement type systems have computational effects, e.g., exception (like division and assertion), divergence, nondeterminism~\cite{unno2018}, and probability~\cite{barthe2015}.
Assume that we have a fibred monad $\hat{T}$ on $p : \category{E} \to \category{B}$, a monad $T$ on $\category{B}$, and a lifting $\dot{T}$ of $T$ along $q : \category{P} \to \category{B}$.
Under a certain condition that roughly claims that $\hat{T}$ and $T$ represent the same computational effects, we construct a fibred monad on $\RefinedCCompC{\category{E}}{\category{P}} \to \category{P}$, which is a lifting of $\hat{T}$ in the same spirit of the given lifting $\dot{T}$.
This situation is rather realistic because the fibred monad $\hat{T}$ on the CCompC $p : \category{E} \to \category{B}$ is often induced from the monad $T$ on the base category $\category{B}$.
The lifting $\dot{T}$ of the monad $T$ along $p : \category{P} \to \category{B}$ specifies how to map predicates $P \in \category{P}_X$ on values $X \in \category{B}$ to predicates $\dot{T} P \in \category{P}_{T X}$ on computations $T X$, which enables us to express, for example, total/partial correctness and may/must nondeterminism~\cite{aguirre2020}.

%Many dependent refinement type systems have computational effects, e.g., partiality (like division and assertion), divergence, nondeterminism~\cite{unno2018}, and probability~\cite{barthe2015}.
%We deal with them by considering liftings of monads~\cite{aguirre2020} along the lifting of CCompCs in Fig.~\ref{fig:lifting_intro}.
%In the presence of computational effects represented by a monad $T$ on $\category{B}$, we need a mapping from predicates on values $X$ to predicates on computations $T X$.
%Liftings $\dot{T}$ of $T$ along the fibration $\category{P} \to \category{B}$ for predicate logic are used for this purpose, and many natural interpretation of predicates on computations, like total/partial correctness and may/must modality, come from liftings.
%We show that under a certain condition, we obtain a fibred monad on $\RefinedCCompC{\category{E}}{\category{P}} \to \category{P}$ from a lifting $\dot{T}$ of $T$ along $q : \category{P} \to \category{B}$ and a fibred monad $\hat{T}$ on $p : \category{E} \to \category{B}$.
%Roughly speaking, this construction of fibred monad liftings means that if $\hat{T}$ on $p$ represents the same computational effects as $T$ on $\category{B}$, then we can lift $\hat{T}$ to a fibred monad on $\RefinedCCompC{\category{E}}{\category{P}} \to \category{P}$ in the same sprit of the lifting $\dot{T}$.
%Moreover, the obtained fibred monad on $\RefinedCCompC{\category{E}}{\category{P}} \to \category{P}$ is a lifting of $\hat{T}$ along the lifting of CCompCs in Fig.~\ref{fig:lifting_intro}.

We explain the usage of these categorical constructions by giving semantics to a dependent refinement type system with computational effects, which is based on~\cite{ahman2016}.
Our system also supports subtyping relations induced by logical implication.
We prove soundness of the refinement type system.

Finally, we discuss how to handle recursion in refinement type systems.
In~\cite{ahman2016}, Ahman gives semantics to recursion in a specific model, i.e., the fibration of continuous families of $\omega$-cpos $\mathbf{CFam}(\CPO) \to \CPO$.
We consider more general characterization of recursion by adapting Conway operators for CCompCs, which enables us to lift the structure for recursion.
We show that a rule for partial correctness in our refinement type system is sound under the existence of a generalized Conway operator.
%However, we find a subtle problem of lifting a Conway operator in $\mathbf{CFam}(\CPO) \to \CPO$ even though $\mathbf{CFam}(\CPO) \to \CPO$ does have a Conway operator.
%So, we leave finding concrete examples of liftings of Conway operators for future work.

%\begin{tikzpicture}
%	\node[draw, align=center] (rts) at (-1.5, 0.8) {refinement \\ type system};
%	\node[draw, align=center] (uts) at (-1.5, -0.8) {underlying \\ type system};
%	\node[draw, align=center] (lifting) at (1.5, 0.8) {lifting of \\ CCompC};
%	\node[draw, align=center] (ccompc) at (1.5, -0.8) {CCompC};
%	\draw[->] (rts) -- (uts);
%\end{tikzpicture}

Our contributions are summarized as follows.
\begin{itemize}
	\item We provide a general construction of liftings of CCompCs from given CCompCs and posetal fibrations satisfying certain conditions, as a semantic counterpart of construction of dependent refinement type systems from underlying type systems and predicate logic.
		We extend this to liftings of fibred monads on the underlying CCompCs to model computational effects.
	\item We consider a type system (based on EMLTT~\cite{ahman2016,ahman2017,ahman2018}) that includes most of basic features of dependent refinement type systems and prove its soundness in the liftings of CCompCs obtained from the above construction.
	\item We define Conway operators for dependent type systems. This generalizes the treatment of general recursion in~\cite{ahman2016}.
		We prove soundness of the typing rule for partial correctness of recursion under the existence of a lifting of Conway operators.
\end{itemize}

\section{Preliminaries}\label{sec:preliminaries}
We review basic definitions and fix notations for comprehension categories, which are used as categorical models for dependent type theories.
We assume basic knowledge of fibrations (see e.g.~\cite{jacobs2001}).

Let $p : \category{E} \to \category{B}$ be a fibration (opfibration).
We denote the cartesian (cocartesian) lifting over $u : I \to J$ by $\CartesianOver{u}{Y} : \pullback{u} Y \to Y$ ($\CoCartesianOver{u}{X} : X \to \coreindexing{u} X$) where $\pullback{u} : \category{E}_J \to \category{E}_I$ ($\coreindexing{u} : \category{E}_I \to \category{E}_J$) is the reindexing (coreindexing) functor.
We call $p : \category{E} \to \category{B}$ a \emph{posetal fibration} if $p$ is a fibration such that each fibre category is a poset.
Note that the fibration $p : \category{E} \to \category{B}$ is split and faithful if $p$ is posetal.

A \emph{comprehension category} is a functor $\functor{P} : \category{E} \to \category{B}^{\to}$ such that the composite $\mathrm{cod} \comp \functor{P} : \category{E} \to \category{B}$ is a fibration and $\functor{P}$ maps cartesian morphisms to pullbacks in $\category{B}$.
A comprehension category $\functor{P}$ is \emph{full} if $\functor{P}$ is fully faithful.

A \emph{comprehension category with unit} is a fibration $p : \category{E} \to \category{B}$ that has a fibred terminal object $1 : \category{B} \to \category{E}$ and a comprehension functor $\ComprehensionFunctor{-} : \category{E} \to \category{B}$ which is a right adjoint of the fibred terminal object functor $1 \dashv \ComprehensionFunctor{-}$.
Projection $\pi_X : \ComprehensionFunctor{X} \to p X$ is defined by $\pi_X = p \epsilon_X^{1 \dashv \ComprehensionFunctor{-}}$ for each $X \in \category{E}$.
Intuitively, $\category{E}$ represents a collection of types $\WellFormedType{\Gamma}{A}$ in dependent type theories; $\category{B}$ represents a collection of contexts $\Gamma$; $p : \category{E} \to \category{B}$ is the mapping $(\WellFormedType{\Gamma}{A}) \mapsto \Gamma$; $1 : \category{B} \to \category{E}$ is the unit type $\Gamma \mapsto (\WellFormedType{\Gamma}{1})$; and $\ComprehensionFunctor{-}$ is the mapping $(\WellFormedType{\Gamma}{A}) \mapsto \Gamma, x : A$ where $x$ is a fresh variable.

The comprehension category with unit $p : \category{E} \to \category{B}$ induces several structures.
It induces a comprehension category $\functor{P}$ defined by $\functor{P} X = \pi_X$.
The adjunction $1 \dashv \ComprehensionFunctor{-}$ defines the bijection $s : \category{E}_I(1 I, X) \cong \{ f : I \to \ComprehensionFunctor{X} \mid \pi_X \comp f = \identity{I} \}$ between vertical morphisms in $\category{E}$ and sections in $\category{B}$.
For each $X, Y \in \category{E}_I$, we have an isomorphism $\phi : \category{E}_{\ComprehensionFunctor{X}}(1 \ComprehensionFunctor{X}, \pullback{\pi_X} Y) \cong \category{E}_{I}(X, Y)$.
Consider the pullback square $\functor{P}(\CartesianOver{\pi_X}{Y})$ where $X, Y \in \category{E}_I$.
%\begin{center}
%	\begin{tikzcd}
%		\ComprehensionFunctor{\pullback{\pi_X} Y} \ar[d, swap, "\pi_{\pullback{\pi_X} Y}"] \ar[r, "\ComprehensionFunctor{\CartesianOver{\pi_X}{Y}}"] \ar[rd, phantom, very near start, "\lrcorner"] & \ComprehensionFunctor{Y} \ar[d, "\pi_Y"] \\
%		\ComprehensionFunctor{X} \ar[r, "\pi_X"] & I
%	\end{tikzcd}
%\end{center}
By the universal property of pullbacks, we have the symmetry isomorphism $\sigma_{X, Y} : \ComprehensionFunctor{\pullback{\pi_X} Y} \to \ComprehensionFunctor{\pullback{\pi_Y} X}$ as a unique morphism $\sigma_{X, Y}$ such that $\pi_{\pullback{\pi_X} Y} = \ComprehensionFunctor{\CartesianOver{\pi_Y}{X}} \comp \sigma_{X, Y}$ and $\ComprehensionFunctor{\CartesianOver{\pi_X}{Y}} = \pi_{\pullback{\pi_Y} X} \comp \sigma_{X, Y}$.
Similarly, we have the diagonal morphism $\delta_X : \ComprehensionFunctor{X} \to \ComprehensionFunctor{\pullback{\pi_X} X}$ as a unique morphism $\delta_X$ such that $\pi_{\pullback{\pi_X} X} \comp \delta_X = \ComprehensionFunctor{\CartesianOver{\pi_X}{X}} \comp \delta_X = \identity{\ComprehensionFunctor{X}}$.

Let $p : \category{E} \to \category{B}$ be a comprehension category with unit and $q : \category{D} \to \category{B}$ be a fibration.
The fibration $q$ has \emph{$p$-products} if $\pullback{\pi_X} : \category{D}_{p X} \to \category{D}_{\ComprehensionFunctor{X}}$ has a right adjoint $\pullback{\pi_X} \dashv \prod_X$ for each $X \in \category{E}$ and these adjunctions satisfy the BC (Beck-Chevalley) condition for each pullback square $\functor{P} f$ where $\functor{P}$ is a comprehension category induced by $p$ and $f$ is a cartesian morphism in $\category{E}$.
Similarly, we define \emph{$p$-coproducts} by $\coprod_X \dashv \pullback{\pi_X}$ and $p$-equality by $\mathrm{Eq}_X \dashv \pullback{\delta_X}$ plus the BC condition for each cartesian morphism (see~\cite[Definition~9.3.5]{jacobs2001} for detail).

A comprehension category with unit $p : \category{E} \to \category{B}$ admits \emph{products} (\emph{coproducts}) if it has $p$-products ($p$-coproducts).
The coproducts are \emph{strong} if the canonical morphism $\kappa : \ComprehensionFunctor{Y} \to \ComprehensionFunctor{\coprod_X Y}$
defined by $\ComprehensionFunctor{\CartesianOver{\pi_X}{\coprod_X Y} \comp \eta^{\pullback{\pi_X} \dashv \coprod_X}}$
is isomorphic for each $X \in \category{E}$ and $Y \in \category{E}_{\ComprehensionFunctor{X}}$.
A \emph{closed comprehension category} (CCompC) is a full comprehension category with unit that admits products and strong coproducts and has a terminal object in the base category.
A \emph{split closed comprehension category} (SCCompC) is a CCompC such that $p$ is a split fibration, and the BC condition for products and coproducts holds strictly (i.e., canonical isomorphisms are identities).
For example, the simple fibration $\simple{\category{B}} : \mathbf{s}(\category{B}) \to \category{B}$ on a cartesian closed category $\category{B}$ is a SCCompC (see~\cite[Theorem~10.5.5]{jacobs2001}).
Another example of SCCompCs is the family fibration $\famset : \mathbf{Fam}(\Set) \to \Set$.
%Products and strong coproducts in CCompCs represents dependent function types $\Pi x {:} A. B$ and dependent pair types $\Sigma x {:} A. B$, respectively.

Fibred coproducts in a comprehension category with unit $p : \category{E} \to \category{B}$ are \emph{strong} if the functor $\langle \pullback{\ComprehensionFunctor{\iota_1}}, \pullback{\ComprehensionFunctor{\iota_2}} \rangle : \category{E}_{\ComprehensionFunctor{X + Y}} \to \category{E}_{\ComprehensionFunctor{X}} \times \category{E}_{\ComprehensionFunctor{Y}}$ is fully faithful where $\iota_1 : X \to X + Y$ and $\iota_2 : Y \to X + Y$ are injections for fibred coproducts.
Strong fibred coproducts are used to interpret coproducts types $A + B$.

\section{Lifting SCCompCs and Fibred Coproducts}\label{sec:lifting_ccompc}
In this section, we give a construction of liftings of SCCompCs with strong fibred coproducts from given SCCompCs with strong fibred coproducts for underlying types and posetal fibrations for predicate logic satisfying appropriate conditions.

\subsection{Lifting SCCompCs}
Let $p : \category{E} \to \category{B}$ be a SCCompC for underlying type systems.
Let $q : \category{P} \to \category{B}$ be a posetal fibration with fibred finite products for predicate logic.
\begin{mydefinition}\label{def:refined_ccompc}
	We define a category $\RefinedCCompC{\category{E}}{\category{B}}$ by the pullback of $q^{\to} : \category{P}^{\to} \to \category{B}^{\to}$ along $\functor{P} : \category{E} \to \category{B}^{\to}$ where the comprehension category $\functor{P}$ is induced by $p : \category{E} \to \category{B}$.
	\begin{center}
		\begin{tikzcd}[row sep=small]
			\RefinedCCompC{\category{E}}{\category{B}} \ar[r, "\pullback{(q^{\to})} \functor{P}"] \ar[d, swap, "\pullback{\functor{P}}(q^{\to})"] \ar[rd, phantom, very near start, "\lrcorner"] & \category{P}^{\to} \ar[d, "q^{\to}"] \\
			\category{E} \ar[r, "\functor{P}"] & \category{B}^{\to}
		\end{tikzcd}
	\end{center}
	That is, objects are tuples $(X, P, Q)$ where $X \in \category{E}$, $P \in \category{P}_{p X}$, $Q \in \category{P}_{\ComprehensionFunctor{X}}$, and $Q \le \pi_X^{*} P$; and morphisms are tuples $(f, g, h) : (X, P, Q) \to (X', P', Q')$ where $f : X \to X'$, $g : P \to P'$, $h : Q \to Q'$, $p f = q g$, and $\ComprehensionFunctor{f} = q h$.
\end{mydefinition}
The intuition of this definition is as follows.
For each object $(X, P, Q) \in \RefinedCCompC{\category{E}}{\category{P}}$, $X$ represents a type $\WellFormedType{\Gamma}{A}$ in the underlying type system, $P$ represents a predicate on the context $\Gamma$, and $Q$ represents the conjunction of a predicate on $\Gamma, v : A$ and the predicate $P$ (thus $Q \le \pi_X^{*} P$ is imposed).
Note that  $\pullback{\functor{P}}(q^{\to}) : \RefinedCCompC{\category{E}}{\category{P}} \to \category{E}$ is faithful because $q$ is faithful.

Let $\refinedccompc{p}{q} : \RefinedCCompC{\category{E}}{\category{P}} \to \category{P}$ be a functor defined by $\mathrm{cod} \comp \pullback{(q^{\to})} \functor{P}$, that is, $(X, P, Q) \mapsto P$.
The functor $\refinedccompc{p}{q}$ inherits (most of) the CCompC structure of $p : \category{E} \to \category{B}$.
\begin{mylemma}
	The functor $\refinedccompc{p}{q} : \RefinedCCompC{\category{E}}{\category{P}} \to \category{P}$ is a split fibration.
	The cartesian lifting of $g : P' \to P$ is given by
	\[ (\CartesianOver{q g}{X}, g, \CartesianOver{\ComprehensionFunctor{\CartesianOver{q g}{X}}}{Q} \comp \pi') : (\pullback{(q g)} X, P', \pullback{\pi_{\pullback{(q g)} X}} P' \land \pullback{\ComprehensionFunctor{\CartesianOver{q g}{X}}} Q) \to (X, P, Q) \]
	where $\pi'$ is a projection for fibred products.
	\qed
\end{mylemma}

\begin{mylemma}\label{lem:lift_fccu_coproduct}
	The fibration $\refinedccompc{p}{q} : \RefinedCCompC{\category{E}}{\category{P}} \to \category{P}$ is a full comprehension category with unit that admits strong coproducts.
\end{mylemma}
\begin{proof}
	The main idea is that the structure in the CCompC $p : \category{E} \to \category{B}$ can be lifted to $\RefinedCCompC{\category{E}}{\category{P}} \to \category{P}$.
	Here, we only show the definition of (object parts of) fibred terminal objects $1 : \category{P} \to \RefinedCCompC{\category{E}}{\category{P}}$, the comprehension functor $\ComprehensionFunctor{-} : \RefinedCCompC{\category{E}}{\category{P}} \to \category{P}$, and coproducts $\coprod_{(X, P, Q)} : \RefinedCCompC{\category{E}}{\category{P}}_{Q} \to \RefinedCCompC{\category{E}}{\category{P}}_{P}$ for each $(X, P, Q) \in \RefinedCCompC{\category{E}}{\category{P}}$.
	\begin{gather*}
		1 P = (1 q P, P, \pullback{\pi_{1 q P}} P) \quad
		\ComprehensionFunctor{(X, P, Q)} = Q \quad
		\coprod_{(X, P, Q)} (Y, Q, R) = (\coprod_X Y, P, \pullback{(\kappa^{-1})} R)
	\end{gather*}
	The rest of the proof is omitted.
	\qed
\end{proof}

The existence of products in $\refinedccompc{p}{q}$ requires additional conditions.
\begin{mylemma}\label{lem:lift_product}
	If $q : \category{P} \to \category{B}$ has fibred exponentials and $p$-products (in addition to fibred finite products), then $\refinedccompc{p}{q} : \RefinedCCompC{\category{E}}{\category{P}} \to \category{P}$ admits products.
\end{mylemma}
\begin{proof}
	We define $\prod_{(X, P, Q)} : \RefinedCCompC{\category{E}}{\category{P}}_{Q} \to \RefinedCCompC{\category{E}}{\category{P}}_{P}$ by
	{\small
	\[ \prod_{(X, P, Q)} (Y, Q, R) = (\prod_X Y, P, \pullback{\pi_{\prod_X Y}} P \land \prod_{\pullback{\pi_{\prod_X Y}} X} \pullback{\sigma_{\prod_X Y, X}} (\pullback{\pi_{\pullback{\pi_X} \prod_X Y}} Q \Rightarrow \pullback{\ComprehensionFunctor{\epsilon^{\pullback{\pi_X} \dashv \prod_X}_Y}} R)). \]
	}
	\begin{center}
		\begin{tikzcd}[column sep=large, row sep=-0.8em]
			Q \in \category{P}_{\ComprehensionFunctor{X}} \ar[rd, pos=0.2, "\pullback{\pi_{\pullback{\pi_X} \prod_X Y}}"] \\
			& \category{P}_{\ComprehensionFunctor{\pullback{\pi_X} \prod_X Y}} \ar[r, "\pullback{\sigma_{\prod_X Y, X}}"] & \category{P}_{\ComprehensionFunctor{\pullback{\pi_{\prod_X Y}} X}} \ar[r, bend left, in=170, out=10, "\prod_{\pullback{\pi_{\prod_X Y}} X}", ""{name=U,below}] & \category{P}_{\ComprehensionFunctor{\prod_X Y}} \ar[l, bend left, in=170, out=10, "\pullback{\pi_{\pullback{\pi_{\prod_X Y}} X}}", ""{name=F, above}] \ar[from=F, to=U, symbol=\dashv] \\
			R \in \category{P}_{\ComprehensionFunctor{Y}} \ar[ru, pos=0.2, swap, "\pullback{\ComprehensionFunctor{\epsilon_Y^{\pullback{\pi_X} \dashv \prod_X}}}"]
		\end{tikzcd}
	\end{center}
	Then, this gives products in $\refinedccompc{p}{q}$ but we omit the lengthy proof.
	\qed
\end{proof}

As a result, we get a lifting of SCCompCs over $p : \category{E} \to \category{B}$. \\
\begin{adjustbox}{valign=C,raise=\strutheight,minipage={1.0\linewidth}}
	\begin{wrapfigure}{r}{9em} % picture on the right
		\vspace{-2em}
		\begin{tikzcd}
			\RefinedCCompC{\category{E}}{\category{P}} \ar[d, "\refinedccompc{p}{q}"] \ar[r, "\pullback{\functor{P}}(q^{\to})"] & \category{E} \ar[d, "p"] \\
			\category{P} \ar[r, "q"] & \category{B}
		\end{tikzcd}
	\end{wrapfigure}% 
	\strut{}
	\vspace*{-0.5cm}
\begin{mytheorem}\label{thm:ccompc4refinement}
	If $p : \category{E} \to \category{B}$ is a SCCompC and $q : \category{P} \to \category{B}$ is a fibred ccc that has $p$-products, then $\refinedccompc{p}{q} : \RefinedCCompC{\category{E}}{\category{P}} \to \category{P}$ is a SCCompC.
	Moreover, $(\pullback{\functor{P}}(q^{\to}), q) : \refinedccompc{p}{q} \to p$ is a morphism of SCCompCs, i.e., a split fibred functor that preserves the CCompC structure strictly.
\end{mytheorem}
\end{adjustbox}
\begin{proof}
	By Lemma~\ref{lem:lift_fccu_coproduct} and Lemma~\ref{lem:lift_product}.
	A terminal object in $\category{P}$ exists because $\category{B}$ has a terminal object and $q : \category{P} \to \category{B}$ has fibred terminal objects.
	It is almost obvious that $(\pullback{\functor{P}}(q^{\to}), q)$ preserves the structure of CCompCs.
	\qed
\end{proof}

\begin{myexample}\label{ex:simple_subobject_ccompc}
	Consider the simple fibration $\simple{\Set} : \mathbf{s}(\Set) \to \Set$ and the subobject fibration $\subobj{\Set} : \mathbf{Sub}(\Set) \to \Set$.
	Objects in $\RefinedCCompC{\mathbf{s}(\Set)}{\mathbf{Sub}(\Set)}$ are tuples $((I, X), P, Q)$ where $(I, X) \in \mathbf{s}(\Set)$, $P \subseteq I$, and $Q \subseteq P \times X \subseteq I \times X$, and morphisms are those in $\mathbf{s}(\Set)$ that preserve predicates.
	In $\refinedccompc{\simple{\Set}}{\subobj{\Set}} : \RefinedCCompC{\mathbf{s}(\Set)}{\mathbf{Sub}(\Set)} \to \mathbf{Sub}(\Set)$, products are given as follows.
	\begin{gather}
		\prod_{((I, X), P, Q)} ((I \times X, Y), Q, R) = \big((I, X \Rightarrow Y), P, \{ (i, f) \in I \times (X \Rightarrow Y) \mid \\ i \in P \land \forall x \in X, (i, x) \in Q \implies ((i, x), f(x)) \in R \} \big) \label{eq:simple_subobject_products}
	\end{gather}
\end{myexample}
\begin{myexample}
	Let $\mathsf{erel} : \mathbf{ERel} \to \Set$ be the fibration of endorelations defined by change-of-base from $\mathbf{Sub}(\Set) \to \Set$ along the functor $X \mapsto X \times X$.
	The fibration $\mathsf{erel}$ is a fibred ccc and has products (i.e.\ right adjoints of  reindexing functors that satisfy the BC condition for each pullback square).
	Therefore, $\mathsf{erel}$ has $p$-products for any comprehension category with unit $p$.
	If we apply Theorem~\ref{thm:ccompc4refinement} to $\mathsf{erel}$ and the simple fibration $\simple{\Set} : \mathbf{s}(\Set) \to \Set$, then products are defined similarly to Example~\ref{ex:simple_subobject_ccompc}.
\end{myexample}
\begin{myexample}
	Consider the family fibration $\famset : \mathbf{Fam}(\Set) \to \Set$ and the subobject fibration $\subobj{\Set} : \mathbf{Sub}(\Set) \to \Set$.
	Objects in $\RefinedCCompC{\mathbf{Fam}(\Set)}{\mathbf{Sub}(\Set)}$ are tuples $((I, X), P, Q)$ where $(I, X) \in \mathbf{Fam}(\Set)$, $P \subseteq I$, and $Q \subseteq \coprod_{i \in P} X i \subseteq \coprod_{i \in I} X i$.
	Note that subsets $Q \subseteq \coprod_{i \in I} X i$ have a one-to-one correspondence with families of subsets $(Q i \subseteq X i)_{i \in I}$ when we define $Q i = \pullback{\iota_i}(Q)$ where $\iota_i : X i \to \coprod_{i \in I} X i$ is the $i$-th injection.
	So, we often identify $Q$ with the family of subsets $Q i \subseteq X i$.
	Products in $\refinedccompc{\famset}{\subobj{\Set}} : \RefinedCCompC{\mathbf{Fam}(\Set)}{\mathbf{Sub}(\Set)} \to \mathbf{Sub}(\Set)$ is defined by modifying~\eqref{eq:simple_subobject_products} for dependent functions.
\end{myexample}

\subsection{Lifting Fibred Comproducts}
A sufficient condition for $\refinedccompc{p}{q} : \RefinedCCompC{\category{E}}{\category{P}} \to \category{P}$ to have strong fibred coproducts is given by the following lemma, which is analogous to \cite[Prop.~4.5.8]{hermida1993}.
\begin{mylemma}\label{lem:fibred_coproduct}
	If
	(1) $p : \category{E} \to \category{B}$ is a CCompC that has strong fibred coproducts
	(2) for each $X, Y \in \category{E}_I$, $X', Y' \in \category{E}_{I'}$, $u : I \to I'$, and pair of cartesian liftings $f : X \to X'$ and $g : Y \to Y'$ over $u$, the following two squares are pullbacks
		\begin{center}
			\begin{tikzcd}[row sep=small]
				\ComprehensionFunctor{X} \ar[rd, very near start, phantom, "\lrcorner"] \ar[r, "\ComprehensionFunctor{\iota_1}"] \ar[d, swap, "\ComprehensionFunctor{f}"] & \ComprehensionFunctor{X + Y} \ar[d, "\ComprehensionFunctor{f + g}"] & \ComprehensionFunctor{Y} \ar[l, swap, "\ComprehensionFunctor{\iota_2}"] \ar[d, "\ComprehensionFunctor{g}"] \ar[ld, very near start, phantom, "\llcorner"] \\
				\ComprehensionFunctor{X'} \ar[r, "\ComprehensionFunctor{\iota_1}"] & \ComprehensionFunctor{X' + Y'} & \ComprehensionFunctor{Y'} \ar[l, swap, "\ComprehensionFunctor{\iota_2}"]
			\end{tikzcd}
		\end{center}
	(3) $q : \category{P} \to \category{B}$ is a fibred distributive category
	(4) for each $X, Y \in \category{E}_I$ and $Z \in \category{E}_{\ComprehensionFunctor{X + Y}}$, $q$ has cocartesian liftings of
		$\ComprehensionFunctor{\iota_1} : \ComprehensionFunctor{X} \to \ComprehensionFunctor{X + Y}$,
		$\ComprehensionFunctor{\iota_2} : \ComprehensionFunctor{Y} \to \ComprehensionFunctor{X + Y}$,
		$\ComprehensionFunctor{\CartesianOver{\ComprehensionFunctor{\iota_1}}{Z}} : \ComprehensionFunctor{\pullback{\ComprehensionFunctor{\iota_1}} Z} \to \ComprehensionFunctor{Z}$, and
		$\ComprehensionFunctor{\CartesianOver{\ComprehensionFunctor{\iota_2}}{Z}} : \ComprehensionFunctor{\pullback{\ComprehensionFunctor{\iota_2}} Z} \to \ComprehensionFunctor{Z}$
		that satisfy the BC condition for each pullback squares and Frobenius,
	then $\refinedccompc{p}{q} : \RefinedCCompC{\category{E}}{\category{P}} \to \category{P}$ has strong fibred coproducts, and the fibred functor $(\pullback{\functor{P}}(q^{\to}), q) : \refinedccompc{p}{q} \to p$ strictly preserves fibred coproducts.
\end{mylemma}
\begin{proof}
	We define fibred coproducts by
	$(X, P, Q) + (Y, P, R) = (X + Y, P, \coreindexing{\ComprehensionFunctor{\iota_1}} Q \lor \coreindexing{\ComprehensionFunctor{\iota_2}} R)$.
	We omit the rest of the proof.
	\qed
\end{proof}
Note that if $q$ is fibred bicartesian closed, then $q$ is a fibred distributive category.

\begin{myexample}
	Consider $\simple{\Set} : \mathbf{s}(\Set) \to \Set$ and $\subobj{\Set} : \mathbf{Sub}(\Set) \to \Set$ (recall Example~\ref{ex:simple_subobject_ccompc}).
	This combination satisfies four conditions in Lemma~\ref{lem:fibred_coproduct}.
	Fibred coproducts in $\RefinedCCompC{\mathbf{s}(\Set)}{\mathbf{Sub}(\Set)} \to \mathbf{Sub}(\Set)$ are defined as follows.
	\[ ((I, X), P, Q) + ((I, Y), P, R) = ((I, X + Y), P, \{ (i, x) \mid (i, x) \in Q \lor (i, x) \in R \}) \]
\end{myexample}

\section{Lifting Monads on SCCompCs}\label{sec:lifting_monad}
Suppose we have a SCCompC $p : \category{E} \to \category{B}$ and a posetal fibration $q : \category{P} \to \category{B}$ as ingredients for $\refinedccompc{p}{q} : \RefinedCCompC{\category{E}}{\category{P}} \to \category{P}$ in Theorem~\ref{thm:ccompc4refinement}.
We explain how to construct a fibred monad on $\refinedccompc{p}{q} : \RefinedCCompC{\category{E}}{\category{P}} \to \category{P}$ from monads on $p$ and $q$.

First, we assume that a monad $T$ on $\category{B}$ and a fibred monad $\hat{T}$ on $p : \category{E} \to \category{B}$ are given.
These monads are intended to represent the same computational effects in underlying type systems, but $T$ is more primitive than $\hat{T}$ (like the maybe monad and the powerset monad on $\Set$), and $\hat{T}$ is induced from $T$ in some natural way (e.g.\ we can define $\hat{T}$ by $(I, X) \mapsto (I, T X)$ on the simple fibration $\mathbf{s}(\Set) \to \Set$).
In such a situation, we often have an oplax monad morphism (Definition~\ref{def:oplax_monad_morphism}) $\theta : \ComprehensionFunctor{\hat{T} ({-})} \to T \ComprehensionFunctor{-}$.
Intuitively, $\theta$ extends the action of $\hat{T}$ on types to contexts, just like strengths of strong monads.
%The fibred monad $\hat{T}$ only acts on types and does not change contexts.
We also need a lifting $\dot{T}$ of $T$ along $q : \category{P} \to \category{B}$ to specify a mapping from predicates on values in $X \in \category{B}$ to predicates on computations in $T X$~\cite{aguirre2020}.
Given all these ingredients and some additional conditions, we define a fibred monad on $\refinedccompc{p}{q} : \RefinedCCompC{\category{E}}{\category{P}} \to \category{P}$, which is a lifting of the fibred monad $\hat{T}$ on $p : \category{E} \to \category{B}$.

\begin{mydefinition}[oplax monad morphism]\label{def:oplax_monad_morphism}
	Let $\category{C}, \category{D}$ be categories, $F : \category{C} \to \category{D}$ be a functor, and $(S, \eta^S, \mu^S)$, $(T, \eta^T, \mu^T)$ be monads on $\category{C}$ and $\category{D}$, respectively.
	A natural transformation $\theta : F S \to T F$ is an \emph{oplax monad morphism} if $\theta$ respects units and multiplications.
	\begin{center}
		\begin{tikzcd}[row sep=small]
			F X \ar[d, swap, "F \eta^S_{X}"] \ar[rd, "\eta^T_{F X}"] \\
			F S X \ar[r, "\theta_X"] & T F X
		\end{tikzcd}\quad
		\begin{tikzcd}[row sep=small]
			F S^2 X \ar[r, "\theta_{S X}"] \ar[d, swap, "F \mu^S_X"] & T F S X \ar[r, "T \theta_X"] & T^2 F X \ar[d, "\mu^T_{F X}"] \\
			F S X \ar[rr, "\theta_X"] & & T F X
		\end{tikzcd}
	\end{center}
\end{mydefinition}

\begin{mytheorem}\label{thm:refined_monad}
	Let $T$ be a monad on $\category{B}$, $\hat{T}$ be a fibred monad on $p : \category{E} \to \category{B}$ (in $\mathbf{Fib}_{\category{B}}$), $\theta : \ComprehensionFunctor{\hat{T} ({-})} \to T\ComprehensionFunctor{-}$ be an oplax monad morphism, and $\dot{T}$ be a fibred lifting of $T$ along $q : \category{P} \to \category{B}$.
	If
	\begin{equation}
		\pullback{\pi_{\hat{T} X}} P \land \pullback{\theta_X} \dot{T} Q \le \pullback{\theta_X} \dot{T}(\pullback{\pi_X} P \land Q) \label{eq:monad_lifting_condition}
	\end{equation}
	holds for each $X \in \category{E}$, $P \in \category{P}_{pX}$ and $Q \in \category{P}_{\ComprehensionFunctor{X}}$, then there exists a fibred monad $S$ on $\refinedccompc{p}{q} : \RefinedCCompC{\category{E}}{\category{P}} \to \category{P}$ such that the fibred functor $\refinedccompc{p}{q} \to p$ in Theorem~\ref{thm:ccompc4refinement} is a fibred monad morphism from $S$ to $\hat{T}$.
\end{mytheorem}
\begin{proof}
	We define $S (X, P, Q) = (\hat{T} X, P, \pullback{\pi_{\hat{T} X}} P \land \pullback{\theta} \dot{T} Q)$.
	Then the monad structure of $\hat{T}$ lifts to $S$.
	The assumption \eqref{eq:monad_lifting_condition} is required to prove that $S$ is fibred.
	\begin{equation}
		\begin{tikzcd}[row sep=tiny,baseline=(bottom.base)]
			\category{P} \ar[d, "q"] & \pullback{\theta} \dot{T} Q \ar[r, "\CartesianOver{\theta}{\dot{T} Q}"] & \dot{T} Q \\
			\category{B} & \ComprehensionFunctor{\hat{T} X} \ar[r, "\theta"] & |[alias=bottom]| T \ComprehensionFunctor{X}
		\end{tikzcd}
		\tag*{\qed}
	\end{equation}
\end{proof}

\begin{myexample}
	Any strong monad $T$ on a CCC $\category{B}$ gives rise to a split fibred monad $\hat{T}$ on the simple fibration $\simple{\category{B}} : \mathbf{s}(\category{B}) \to \category{B}$ (actually, there is a one-to-one correspondence~\cite[Ex.2.6.10]{jacobs2001}).
	The monad $\hat{T}$ is defined by $(I, X) \mapsto (I, T X)$.
	An oplax monad morphism $\theta : I \times T X \to T (I \times X)$ is given by the strength.

	Now consider the case where $\category{B} = \Set$.
	Since the strength for the monad $T$ on $\Set$ is given uniquely~\cite[Proposition~3.4]{moggi1991}, we can prove that~\eqref{eq:monad_lifting_condition} holds for any fibred lifting of $T$ along the subobject fibration $\subobj{\Set} : \mathbf{Sub}(\Set) \to \Set$.

	Let $T$ be the maybe monad $({-}) + \{ * \}$.
	There are two fibred liftings of $T$:
	\begin{equation}
		\dot{T}_1 (P \subseteq I) = (P + \{ * \} \subseteq I + \{ * \}) \qquad
		\dot{T}_2 (P \subseteq I) = (P \subseteq I + \{ * \})
	\end{equation}
	for each $(P \subseteq I) \in \mathbf{Sub}(\Set)$.
	The lifting $\dot{T}_1$ corresponds to partial correctness, and $\dot{T}_2$ corresponds to total correctness.
	The fibred monads on $\refinedccompc{\simple{\Set}}{\subobj{\Set}}$ defined in Theorem~\ref{thm:refined_monad} from $\dot{T}_1$ and $\dot{T}_2$ are given by
	\begin{align}
		((I, X), P, Q) &\mapsto \big((I, X + \{ * \}), P, \{ (i, x) \mid (i \in P \land x = *) \lor (i, x) \in Q \}\big) \\
		((I, X), P, Q) &\mapsto \big((I, X + \{ * \}), P, \{ (i, x) \mid (i, x) \in Q \}\big)
	\end{align}
	respectively.
	Here, we leave the left/right injection of coproducts implicit.
\end{myexample}

\begin{myexample}
	For each monad $T$ on $\Set$, we have a split fibred monad on the family fibration $\mathbf{Fam}(\Set) \to \Set$ defined by $\hat{T} (I, X) = (I, T \comp X)$.
	We have an oplax monad morphism $\theta : \coprod_{i \in I} T X i \to T \coprod_{i \in I} X i$ defined by the cotupling $[(T \iota_i)_{i \in I}] : \coprod_{i \in I} T X i \to T \coprod_{i \in I} X i$ where $\iota_i : X i \to \coprod_{i \in I} X i$ is the $i$-th injection.
	The condition~\eqref{eq:monad_lifting_condition} holds for any fibred lifting of $T$ along the subobject fibration $\mathbf{Sub}(\Set) \to \Set$.
	Moreover, we have $\pullback{\iota_i} \pullback{\theta} \dot{T} Q = \dot{T} \pullback{\iota_i} Q$ for each $Q \in \mathbf{Sub}(\Set)_{\coprod_{i \in I} X i}$, so the monad in Theorem~\ref{thm:refined_monad} is given by
	\[ \big((I, X), P, (Q i \subseteq X i)_{i \in I}\big) \mapsto \big((I, T \comp X), P, (\dot{T} Q i \subseteq T X i)_{i \in I}\big). \]
\end{myexample}

\section{Soundness}\label{sec:type_system}
We consider a concrete refinement type system with computational effects and define sound semantics to show that the SCCompC defined in Theorem~\ref{thm:ccompc4refinement} has sufficient structures for dependent refinement types.
Here, we consider two type systems.
One is an underlying type system that is a fragment of EMLTT~\cite{ahman2016,ahman2017,ahman2018}.
The other is a refinement of the underlying type system that has refinement types $\{ v : A \mid p \}$ and a subtyping relation $\Subtyping{\Gamma}{A}{B}$ induced by logical implication.
The two type systems share a common syntax for terms while types are more expressive in the refinement type system.
We consider liftings of fibred adjunction models to interpret the refinement type system.
Here, Theorem~\ref{thm:refined_monad} can be used to obtain a lifting of fibred adjunction models via Eilenberg-Moore construction.
We prove a soundness theorem that claims if a term is well-typed in the refinement type system, then the interpretation of the term has a lifting along the morphism of CCompCs defined in Theorem~\ref{thm:ccompc4refinement}.

\subsection{Underlying Type System}\label{subsec:underlying_type_system}
We define the underlying dependent type system by a slightly modified version of a fragment of EMLTT~\cite{ahman2016,ahman2017,ahman2018}.
We remove some of the types and terms from the original for simplicity.
We parameterize our type system with a set of base type constructors (ranged over by $b$) and a set of value constants (ranged over by $c$) for convenience.

We define value types ($A, B, \dots$), computation types ($\underline{C}, \underline{D}, \dots$), contexts ($\Gamma, \dots$), value terms ($V, W, \dots$), and computation terms ($M, N, \dots$) as follows.
\begin{align}
	A &\coloneqq 1 \mid b_{A}(V) \mid \ValueSumType{x}{A}{B} \mid U \underline{C} \mid A + B \\
	\underline{C} &\coloneqq F A \mid \CompProdType{x}{A}{\underline{C}} \qquad\qquad\qquad
	\Gamma \coloneqq \emptyctx \mid \Gamma, x : A \\
	V &\coloneqq x \mid * \mid c_{A} \mid \langle V, W \rangle_{(x : A). B} \mid \mathbf{thunk}\ M \mid \mathbf{inl}_{A+B}\ V \mid \mathbf{inr}_{A+B}\ V \\
	M &\coloneqq \mathbf{return}\ V \mid M\ \mathbf{to}\ x : A\ \mathbf{in}_{\underline{C}}\ N \mid \mathbf{force}_{\underline{C}}\ V \mid \lambda x : A. M \mid M(V)_{(x : A). \underline{C}} \mid \\
	&\qquad \PatternMatch{V}{x}{A}{y}{B}{z}{\underline{C}}{M} \mid \\
	&\qquad \Case{V}{z}{\underline{C}}{x}{A}{M}{y}{B}{N}
\end{align}
We implicitly assume that variables in $\Gamma$ are mutually different.
We use many type annotations in the syntax of terms for a technical reason, but we might omit them if they are clear from the context.
We define substitution $A[V/x]$, $\underline{C}[V/x]$, $W[V/x]$, and $M[V/x]$ as usual.

For each type constructor $b$, let $\mathrm{arg}(b)$ be a closed value type of the argument of $b$.
We write $b : A \to \mathrm{Type}$ if $A = \mathrm{arg}(b)$.
For each value constant $c$, let $\mathrm{ty}(c)$ be a closed value type of $c$.

We have several kinds of judgements: well-formed contexts $\WellFormedContext{\Gamma}$; well-formed (value or computation) types $\WellFormedType{\Gamma}{A}$, $\WellFormedType{\Gamma}{\underline{C}}$; well-typed (value or computation) terms $\WellTypedTerm{\Gamma}{V}{A}$, $\WellTypedTerm{\Gamma}{M}{\underline{C}}$; and definitional equalities for contexts, types and terms $\ContextEqual{\Gamma_1}{\Gamma_2}$, $\TypeEqual{\Gamma}{A}{B}$, $\TypeEqual{\Gamma}{\underline{C}}{\underline{D}}$, $\TermEqual{\Gamma}{V}{W}{A}$, $\TermEqual{\Gamma}{M}{N}{\underline{C}}$.

Typing rules are basically the same as EMLTT.
Rules for base type constructors and value constants are shown in Fig.~\ref{fig:typing_underlying}

\begin{figure}[tb]
	\begin{mathpar}
		\inferrule{
			\WellFormedContext{\Gamma} \\
			\WellFormedType{\emptyctx}{\mathrm{ty}(c)}
		}{\WellTypedTerm{\Gamma}{c_{\mathrm{ty}(c)}}{\mathrm{ty}(c)}}
		\and
		\inferrule{
			b : A \to \mathrm{Type} \\\\
			\WellFormedType{\emptyctx}{A} \\
			\WellTypedTerm{\Gamma}{V}{A}
		}{\WellFormedType{\Gamma}{b_A(V)}}
		\and
		\inferrule{
			b : A \to \mathrm{Type} \\
			\WellFormedType{\emptyctx}{A} \\\\
			\TermEqual{\Gamma}{V}{W}{A}
		}{\TypeEqual{\Gamma}{b_A(V)}{b_A(W)}}
	\end{mathpar}
	\caption{Some typing rules for the underlying type system.}
	\label{fig:typing_underlying}
\end{figure}

\paragraph{Semantics.}
We use fibred adjunction models to interpret terms and types.
We adapt the definition for our fragment of EMLTT as follows.
\begin{mydefinition}[Fibred adjunction models]\label{def:fibred_adjunction_model}
	A \emph{fibred adjunction model} is a fibred adjunction $F \dashv U : r \to p$ where $p : \category{V} \to \category{B}$ is a SCCompC with strong fibred coproducts and $r : \category{C} \to \category{B}$ is a fibration with $p$-products.
\end{mydefinition}

The Eilenberg-Moore fibration of a CCompC $p : \category{E} \to \category{B}$ inherits products in $p$~\cite[Theorem~4.3.24]{ahman2017} and thus gives an example of fibred adjunction models.
\begin{mylemma}\label{lem:eilenberg_moore_fibred_adjunction_model}
	Given a SCCompC $p : \category{E} \to \category{B}$ with strong fibred products and a split fibred monad $T$ on $p$, then the Eilenberg-Moore adjunction of $T$ is a fibred adjunction model.
	\qed
\end{mylemma}

We assume that a fibred adjunction model $F \dashv U : r \to p$ between $p : \category{E} \to \category{B}$ and $r : \category{C} \to \category{B}$ is given and that interpretations of base type constructors $\llbracket b \rrbracket \in \category{E}$ and value constants $\llbracket c \rrbracket \in \category{E}_1(1, X)$ (for some $X \in \category{E}_1$) are given.
We define a partial interpretation $\llbracket {-} \rrbracket$ of the following form for raw syntax.
\begin{center}
	\begin{tikzcd}[column sep=tiny]
		\category{E} \ar[rd, swap, "p"] \ar[rr, bend left, "F", ""{name=F}] & & \category{C} \ar[ld, "r"] \ar[ll, bend left, "U", ""{name=U}] \ar[from=F, to=U, symbol=\dashv] \\
		& \category{B}
	\end{tikzcd}
	\begin{minipage}{24em}
		\vspace*{-2em}
		\begin{gather*}
			\llbracket \Gamma \rrbracket \in \category{B} \qquad
			\llbracket \Gamma; A \rrbracket \in \category{E}_{\llbracket \Gamma \rrbracket} \qquad
			\llbracket \Gamma; \underline{C} \rrbracket \in \category{C}_{\llbracket \Gamma \rrbracket} \\
			\llbracket \Gamma; V \rrbracket \in \category{E}_{\llbracket \Gamma \rrbracket}(1 \llbracket \Gamma \rrbracket, A) \qquad\text{for some $A$} \\
			\llbracket \Gamma; M \rrbracket \in \category{E}_{\llbracket \Gamma \rrbracket}(1 \llbracket \Gamma \rrbracket, U C) \qquad\text{for some $C \in \category{C}$}
		\end{gather*}
	\end{minipage}
\end{center}
Most of the definition of $\llbracket {-} \rrbracket$ are the same as~\cite{ahman2017}.
For base type constructors $b$ and value constants $c$, we define $\llbracket {-} \rrbracket$ as follows.
\begin{gather}
	\llbracket \Gamma; b_A(V) \rrbracket = \pullback{(s \llbracket \Gamma; V \rrbracket)} \pullback{\ComprehensionFunctor{\CartesianOver{!_{\llbracket \Gamma \rrbracket}}{\llbracket \emptyctx; A \rrbracket}}} \llbracket b \rrbracket \qquad
	\llbracket \Gamma; c_A \rrbracket = \pullback{{!}_{\llbracket \Gamma \rrbracket}} \llbracket c \rrbracket
\end{gather}
Here, left-hand sides are defined if right-hand sides are defined.

\begin{myproposition}[Soundness]\label{prop:soundness_underlying}
	Assume that $\llbracket b \rrbracket \in \category{E}_{\ComprehensionFunctor{\llbracket \emptyctx; A \rrbracket}}$ holds for each $b : A \to \mathrm{Type}$ such that $\llbracket \emptyctx; A \rrbracket$ is defined, and $\llbracket c \rrbracket \in \category{E}_1(1, \llbracket \emptyctx; \mathrm{ty}(c) \rrbracket)$ holds if $\llbracket \emptyctx; \mathrm{ty}(c) \rrbracket \allowbreak \in \category{E}_1$ is defined.
	Interpretations $\llbracket {-} \rrbracket$ of well-formed contexts and types and well-typed terms are defined.
	If two contexts, types, or terms are definitionally equal, then their interpretations are equal.
	\qed
\end{myproposition}

\subsection{Predicate Logic}\label{subsec:predicate_logic}
We define syntax for logical formulas by
\[ p = \top \mid p \land q \mid p \Rightarrow q \mid \forall x : A. p \mid V =_A W \mid a(V) \]
where $a$ ranges over predicate symbols.
Here, we added $\top$ and $V =_A W$ for typing rule for the unique value of the unit type and variables of base types (i.e.\ for selfification), respectively, which we describe later.
However, there is a large amount of freedom to choose the syntax of logical formulas.
The least requirement here is that logical formulas can be interpreted in a posetal fibration $q : \category{P} \to \category{B}$, and interpretations of logical formulas admit semantic weakening, substitution, and conversion in the sense of~\cite[Proposition~5.2.4,~5.2.6]{ahman2017}.
So, we can almost freely add or remove logical connectives and quantifiers as long as $q : \category{P} \to \category{B}$ admits them.

We define a standard judgement of well-formedness for logical formulas.
Some of the rules for well-formedness are shown in Fig.~\ref{fig:well_formed_predicate}
\begin{figure}[tb]
	\begin{mathpar}
		\inferrule{
			\WellTypedTerm{\Gamma}{V}{A} \\
			\WellTypedTerm{\Gamma}{W}{A}
		}{\Gamma \vdash V =_A W : \mathrm{Prop}}
		\and
		\inferrule{
			a : A \to \mathrm{Prop} \\
			\WellFormedType{\emptyctx}{A} \\
			\WellTypedTerm{\Gamma}{V}{A}
		}{\Gamma \vdash a(V) : \mathrm{Prop}}
	\end{mathpar}
	\caption{Some rules for well-formed predicates.}
	\label{fig:well_formed_predicate}
\end{figure}

Logical formulas are interpreted in the fibration $q : \category{P} \to \category{B}$.
We assume that interpretation $\llbracket a \rrbracket \in \category{P}_{\ComprehensionFunctor{\llbracket \emptyctx; A \rrbracket}}$ for each predicate symbol $a : A \to \mathrm{Prop}$ is given.
The interpretation $\llbracket \Gamma \vdash p \rrbracket \in \category{P}_{\llbracket \Gamma \rrbracket}$ is standard and defined inductively for each well-formed formulas.
For example:
\begin{align*}
	\llbracket \Gamma \vdash V =_A W \rrbracket &= \pullback{(s \llbracket \Gamma; V \rrbracket)} \pullback{(s(\pullback{\pi_{\llbracket \Gamma; A \rrbracket}} \llbracket \Gamma; W \rrbracket))} \mathrm{Eq}(\top \ComprehensionFunctor{\llbracket \Gamma; A \rrbracket}) \\
	\llbracket \Gamma \vdash a(V) \rrbracket &= \pullback{s(\llbracket \Gamma; V \rrbracket)} \pullback{\ComprehensionFunctor{\CartesianOver{!_{\llbracket \Gamma \rrbracket}}{\llbracket \emptyctx; A \rrbracket}}} \llbracket a \rrbracket
\end{align*}
where $a : A \to \mathrm{Prop}$ is a predicate symbol.

\subsection{Refinement Type System}
We refine the underlying type system by adding predicates to base types and the unit type.
From now on, we use subscript $A_u$ for types in the underlying type system to distinguish them from types in the refinement type system.
\begin{align}
	A &\coloneqq \{ v : b_{\underlying{A}}(V) \mid p \} \bigm| \{ v : 1 \mid p \} \bigm| \ValueSumType{x}{A}{B} \bigm| U \underline{C} \mid A + B \\
	\underline{C} &\coloneqq F A \mid \CompProdType{x}{A}{\underline{C}} \qquad\qquad\qquad
	\Gamma \coloneqq \emptyctx \mid \Gamma, x : A
\end{align}
We use the same definition of terms as the underlying type system and the same set of base type constructors and value constants.
Argument types of base type constructors $b : A_u \to \mathrm{Type}$ are also the same, but types $\mathrm{ty}(c)$ assigned to value constants $c$ are redefined as refinement types.
Given a type $A$ (or $\underline{C}$) in the refinement type system, we define its underlying type $\ElimRefinement{A}$ (or $\ElimRefinement{\underline{C}}$) by induction where predicates are eliminated in the base cases.
\begin{gather}
	\ElimRefinement{\{ v : b_{A_u}(V) \mid p \}} = b_{A_u}(V) \qquad
	\ElimRefinement{\{ v : 1 \mid p \}} = 1
\end{gather}
Underlying contexts $\ElimRefinement{\Gamma}$ are also defined by $\ElimRefinement{\emptyctx} = \emptyctx$ and $\ElimRefinement{\Gamma, x : A} = \ElimRefinement{\Gamma}, x : \ElimRefinement{A}$.

Judgements in the refinement type system are as follows.
We have judgements for well-formedness or well-typedness for contexts, types and terms in the refinement type system, which are denoted in the same way as the underlying type system.
We do not consider definitional equalities for terms because they are the same as the underlying type system.
Instead, we add judgements for subtyping between types and contexts.
They are denoted by $\ContextSubtyping{\Gamma_1}{\Gamma_2}$ for context, $\Subtyping{\Gamma}{A}{B}$ for value types, and $\Subtyping{\Gamma}{\underline{C}}{\underline{D}}$ for computation types.

Most of term and type formation rules are similar to the underlying type system.
We listed some of the non-trivial modifications of typing rules in Fig.~\ref{fig:typing_refinement}.
We add typing rules for $\{ v : b_{B_u}(V) \mid p \}$ and $\{ v : 1 \mid p \}$.
Subtyping for these types are defined by judgements $\Gamma; v : A_u \mid p \vdash q$ for logical implication.
Here,  $\Gamma; v : A_u \mid p \vdash q$ means ``assumptions in $\Gamma$ and $p$ implies $q$'' where $p$ and $q$ are well-formed formulas in the context $\ElimRefinement{\Gamma}, v : A_u$.
%Note that $\Gamma$ is a context in the refinement type system while $A_u$ is a type in the underlying type system.
We do not specify derivation rules for the judgement $\Gamma; v : A_u \mid p \vdash q$ but assume soundness of the judgement (explained later).
We allow ``selfification''~\cite{ou2004} for variables of base types.
Subtyping for $\ValueSumType{x}{A}{B}$, $U \underline{C}$, $F A$, and $\CompProdType{x}{A}{\underline{C}}$ are defined covariantly except the argument type $A$ of $\CompProdType{x}{A}{\underline{C}}$, which is contravariant.
We have the rule of subsumption.
Value constants are typed with a refined type assignment $\mathrm{ty}(c)$.
The unique value $*$ of the unit type has type $\{ v : 1 \mid \top \}$.

\begin{figure}[tb]
	\begin{mathpar}
		\inferrule{
			b : \underlying{A} \to \mathrm{Type} \\
			\WellFormedContext{\Gamma} \\
			\WellFormedType{\ElimRefinement{\Gamma}}{b_{A_u}(V)} \\\\
			\ElimRefinement{\Gamma}, v : b_{A_u}(V) \vdash p : \mathrm{Prop}
		}{\WellFormedType{\Gamma}{\{ v : b_{A_u}(V) \mid p \}}}
		\and
		\inferrule{
			\WellFormedContext{\Gamma} \\
			\TypeEqual{\ElimRefinement{\Gamma}}{b_{A_u}(V)}{b_{A_u}(W)} \\\\
			\Gamma; v : b_{A_u}(V) \mid p \vdash q
		}{\Subtyping{\Gamma}{\{ v : b_{A_u}(V) \mid p \}}{\{ v : b_{A_u}(W) \mid q \}}}
		\and
		\inferrule{\WellFormedContext{\Gamma_1, x : \{ v : b_{A_u}(V) \mid p \}, \Gamma_2}}{\WellTypedTerm{\Gamma_1, x : \{ v : b_{A_u}(V) \mid p \}, \Gamma_2}{x}{\{ v : b_{A_u}(V) \mid v = x \}}}
		\and
		\inferrule{
			\WellFormedContext{\Gamma} \\
			\WellFormedType{\emptyctx}{\mathrm{ty}(c)}
		}{\WellTypedTerm{\Gamma}{c_{\ElimRefinement{\mathrm{ty}(c)}}}{\mathrm{ty}(c)}}
		\and
		\inferrule{
			\Subtyping{\Gamma}{A_2}{A_1} \\\\
			\WellFormedType{\Gamma, x : A_1}{\underline{C}_1} \\
			\Subtyping{\Gamma, x : A_2}{\underline{C}_1}{\underline{C}_2}
		}{\Subtyping{\Gamma}{\CompProdType{x}{A_1}{\underline{C}_1}}{\CompProdType{x}{A_2}{\underline{C}_2}}}
		\and
		\inferrule{
			\WellTypedTerm{\Gamma_2}{V}{A} \\\\	
			\ContextSubtyping{\Gamma_1}{\Gamma_2} \\
			\Subtyping{\Gamma_1}{A}{B}
		}{\WellTypedTerm{\Gamma_1}{V}{B}}
		\and
		\inferrule{\WellFormedContext{\Gamma}}{\WellTypedTerm{\Gamma}{*}{\{ v : 1 \mid \top \}}}
		\and
		\inferrule{
			\WellFormedContext{\Gamma} \\
			\ElimRefinement{\Gamma}, v : 1 \vdash p : \mathrm{Prop}
		}{\WellFormedType{\Gamma}{\{ v : 1 \mid p \}}}
		\and
		\inferrule{
			\WellFormedContext{\Gamma} \\
			\Gamma; v : 1 \mid p \vdash q
		}{\Subtyping{\Gamma}{\{ v : 1 \mid p \}}{\{ v : 1 \mid q \}}}
	\end{mathpar}
	\caption{Some typing rules for the refinement type system.}
	\label{fig:typing_refinement}
\end{figure}
\begin{mylemma}\label{lem:refinement_underlying_type_system}
	If we eliminate predicates in the refinement types from well-formed contexts, types and terms, then we get well-formed contexts, types and terms of the underlying type system.
	\begin{itemize}
		\item If $\WellFormedContext{\Gamma}$, then $\WellFormedContext{\ElimRefinement{\Gamma}}$.
			If $\WellFormedType{\Gamma}{A}$, then $\WellFormedType{\ElimRefinement{\Gamma}}{\ElimRefinement{A}}$.
			If $\WellFormedType{\Gamma}{\underline{C}}$, then $\WellFormedType{\ElimRefinement{\Gamma}}{\ElimRefinement{\underline{C}}}$.
		\item If $\ContextSubtyping{\Gamma_1}{\Gamma_2}$, then $\ContextEqual{\ElimRefinement{\Gamma_1}}{\ElimRefinement{\Gamma_2}}$.
			If $\Subtyping{\Gamma}{A}{B}$, then $\TypeEqual{\ElimRefinement{\Gamma}}{\ElimRefinement{A}}{\ElimRefinement{B}}$.
			If $\Subtyping{\Gamma}{\underline{C}}{\underline{D}}$, then $\TypeEqual{\ElimRefinement{\Gamma}}{\ElimRefinement{\underline{C}}}{\ElimRefinement{\underline{D}}}$.
	\end{itemize}
\end{mylemma}
\begin{proof}
	By induction on the derivation of judgements.
	Each typing rule in the refinement type system has a corresponding rule in the underlying system.
	\qed
\end{proof}

\begin{myexample}
	We can express conditional branching using the elimination rule of the fibred coproduct type $1 + 1$.
	For example, assume we have a base type constructor $\mathrm{int} : 1 \to \mathrm{Type}$ for integers and a value constant for comparison.
	\[ ({\le}) : U (\CompProdType{x}{\mathrm{int}}{\CompProdType{y}{\mathrm{int}}{F(\{ v : 1 \mid x \le y \} + \{ v : 1 \mid x > y \})}}) \]
	We can define $\mathbf{if}\ x \le y\ \mathbf{then}\ M\ \mathbf{else}\ N$ to be a syntax sugar for
	\[ (x \le' y)\ \mathbf{to}\ z\ \mathbf{in}\ (\mathbf{case}\ z\ \mathbf{of}\ (\mathbf{inl}\ v \mapsto M, \mathbf{inr}\ v \mapsto N)) \]
	where $({\le'}) = \force{}{({\le})}$.
	Note that $M$ and $N$ are typed in contexts that have $v : \{ v : 1 \mid x \le y \}$ or $v : \{ v : 1 \mid x > y \}$ depending on the result of comparison.
\end{myexample}

%\begin{myexample}
%	cbn/cbv
%\end{myexample}

\subsection{Semantics}
\begin{mydefinition}[lifting of fibred adjunction models]
	Suppose that we have two fibred adjunction models
	$F \dashv U : q \to p$ between $p : \category{V} \to \category{B}$ and $q : \category{C} \to \category{B}$ and $\dot{F} \dashv \dot{U} : s \to r$ between $r : \category{U} \to \category{P}$ and $s : \category{D} \to \category{P}$.
	The fibred adjunction model $\dot{F} \dashv \dot{U}$ is a \emph{lifting} of $F \dashv U$ if there exists functors $u : \category{U} \to \category{V}$, $v : \category{D} \to \category{C}$, and $t : \category{P} \to \category{B}$ such that these functors strictly preserve all structures of $\dot{F} \dashv \dot{U}$ to those of $F \dashv U$.
	That is, $(u, t) : r \to p$ and $(v, t) : s \to q$ are split fibred functors,
	the pair of fibred functor $(u, t)$ and $(v, t)$ is a map of adjunctions in $\mathbf{Fib}$,
	$(u, t)$ strictly preserves the CCompC structure and fibred coproducts, and
	$(v, t)$ maps $r$-products to $p$-products in the strict sense.
%	\begin{center}
%		\begin{tikzcd}[column sep=tiny]
%			\category{V} \ar[rd, swap, "p"] \ar[rr, bend left, "F", ""{name=F}] & & \category{C} \ar[ld, "q"] \ar[ll, bend left, "U", ""{name=U}] \ar[from=F, to=U, symbol=\dashv] \\
%			& \category{B}
%		\end{tikzcd}
%		\begin{tikzcd}[column sep=tiny]
%			\category{U} \ar[rd, swap, "r"] \ar[rr, bend left, "\dot{F}", ""{name=F}] & & \category{D} \ar[ld, "s"] \ar[ll, bend left, "\dot{U}", ""{name=U}] \ar[from=F, to=U, symbol=\dashv] \\
%			& \category{P}
%		\end{tikzcd}
%		\begin{tikzcd}
%			\category{U} \ar[r, "u"] \ar[d, "r"] & \category{V} \ar[d, "p"] \\
%			\category{P} \ar[r, "t"] & \category{B}
%		\end{tikzcd}
%		\begin{tikzcd}
%			\category{D} \ar[r, "v"] \ar[d, "s"] & \category{C} \ar[d, "q"] \\
%			\category{P} \ar[r, "t"] & \category{B}
%		\end{tikzcd}
%	\end{center}
\end{mydefinition}

%\begin{mylemma}
%	Suppose we have two SCCompCs $p : \category{D} \to \category{A}$ and $q : \category{E} \to \category{B}$ with strong fibred coproducts, a morphism of CCompC $(u, v) : p \to q$ that strictly preserves fibred coproducts, a split fibred monad $S$ on $p$, a split fibred monad $T$ on $q$.
%	Assume that $(u, v)$ is a split fibred monad morphism from $S$ to $T$.
%	The Eilenberg-Moore adjunction for $S$ gives a lifting of fibred adjunction models over the Eilenberg-Moore adjunction for $T$.
%	\qed
%\end{mylemma}

We assume that a lifting of fibred adjunction models is given as follows.
\begin{equation}
	\begin{tikzcd}[column sep=tiny, row sep=small]
		\category{E} \ar[rd, swap, "p"] \ar[rr, bend left, in=160, out=20, "F", ""{name=F}] & & \category{C} \ar[ld] \ar[ll, bend left, in=160, out=20, "U", ""{name=U}] \ar[from=F, to=U, symbol=\dashv] \\
		& \category{B}
	\end{tikzcd}\quad
	\begin{tikzcd}[column sep=tiny, row sep=small]
		\RefinedCCompC{\category{E}}{\category{P}} \ar[rd, swap, "\refinedccompc{p}{q}"] \ar[rr, bend left, in=170, out=10, "\dot{F}", ""{name=F}, pos=0.3] & & \category{D} \ar[ld] \ar[ll, bend left, in=170, out=10, "\dot{U}", ""{name=U}, pos=0.7] \ar[from=F, to=U, symbol=\dashv] \\
		& \category{P}
	\end{tikzcd}\quad
	\begin{tikzcd}[row sep=small]
		\RefinedCCompC{\category{E}}{\category{P}} \ar[r, "u"] \ar[d, "\refinedccompc{p}{q}"] & \category{E} \ar[d, "p"] \\
		\category{P} \ar[r, "q"] & \category{B}
	\end{tikzcd}\quad
	\begin{tikzcd}[row sep=small]
		\category{D} \ar[r, "v"] \ar[d] & \category{C} \ar[d] \\
		\category{P} \ar[r, "q"] & \category{B}
	\end{tikzcd}
	\label{eq:lifting_fibred_adjunction_model}
\end{equation}
Here, we assume more than just a lifting of fibred adjunction models by requiring the specific SCCompC $\refinedccompc{p}{q}$ with strong fibred coproducts, and the split functor $(u, q) : \refinedccompc{p}{q} \to p$ defined in Theorem~\ref{thm:ccompc4refinement} and Lemma~\ref{lem:fibred_coproduct}.
The underlying fibred adjunction model $F \dashv U$ is used for the underlying type system in \S\ref{subsec:underlying_type_system}, and $q : \category{P} \to \category{B}$ is for predicate logic in \S\ref{subsec:predicate_logic}.
One way to obtain such liftings of fibred adjunction models is to apply the Eilenberg-Moore construction to the monad morphism in Theorem~\ref{thm:refined_monad}, but in general we do not restrict $\category{C}$ and $\category{D}$ to be Eilenberg-Moore categories.
We further assume that $q$ has $p$-equalities to interpret logical formulas of the form $V =_A W$.

We define partial interpretation of refinement types $\llbracket \Gamma \rrbracket \in \category{P}$, $\llbracket \Gamma; A \rrbracket \in \RefinedCCompC{\category{E}}{\category{P}}_{\llbracket \Gamma \rrbracket}$, and $\llbracket \Gamma; \underline{C} \rrbracket \in \category{D}_{\llbracket \Gamma \rrbracket}$ similarly to the underlying type system but with the following modification.
Here, we make use of the definition of $\RefinedCCompC{\category{E}}{\category{P}}$.
\begin{align}
	\llbracket \Gamma; \{ v : b(V) \mid p \} \rrbracket &= \big(\llbracket \ElimRefinement{\Gamma}; b(V) \rrbracket, \llbracket \Gamma \rrbracket, \pullback{\pi_{\llbracket \ElimRefinement{\Gamma}; b(V) \rrbracket}} \llbracket \Gamma \rrbracket \land \llbracket \ElimRefinement{\Gamma}, v : b(V) \vdash p \rrbracket\big) \\
	\llbracket \Gamma; \{ v : 1 \mid p \} \rrbracket &= \big(\llbracket \ElimRefinement{\Gamma}; 1 \rrbracket, \llbracket \Gamma \rrbracket, \pullback{\pi_{\llbracket \ElimRefinement{\Gamma}; 1 \rrbracket}} \llbracket \Gamma \rrbracket \land \llbracket \ElimRefinement{\Gamma}, v : 1 \vdash p \rrbracket\big)
\end{align}
For each $(X, P, Q), (X', P', Q') \in \RefinedCCompC{\category{E}}{\category{P}}$, we define a semantic subtyping relation $(X, P, Q) \subtype (X', P', Q')$ by the conjunction of $X = X'$, $P = P'$, and $Q \le Q'$.
In other words, we have $(X, P, Q) \subtype (X', P', Q')$ if and only if there exists a morphism $(\identity{X}, \identity{P}, h) : (X, P, Q) \to (X', P', Q')$ that is mapped to identities by $u : \RefinedCCompC{\category{E}}{\category{P}} \to \category{E}$ and $\refinedccompc{p}{q} : \RefinedCCompC{\category{E}}{\category{P}} \to \category{P}$.

\begin{mylemma}\label{lem:denotation_refine_underlying}
	\begin{itemize}
		\item If $\llbracket \Gamma \rrbracket$ is defined, then $\llbracket \ElimRefinement{\Gamma} \rrbracket$ is defined and equal to $q \llbracket \Gamma \rrbracket$.
		\item If $\llbracket \Gamma; A \rrbracket$ is defined, then $\llbracket \ElimRefinement{\Gamma}; \ElimRefinement{A} \rrbracket$ is defined and equal to $u \llbracket \Gamma; A \rrbracket$.
		\item If $\llbracket \Gamma; \underline{C} \rrbracket$ is defined, then $\llbracket \ElimRefinement{\Gamma}; \ElimRefinement{\underline{C}} \rrbracket$ is defined and equal to $v \llbracket \Gamma; \underline{C} \rrbracket$.
	\end{itemize}
\end{mylemma}
\begin{proof}
	By simultaneous induction.
	The case of $\{ v : A_u \mid p \}$ is obvious, and other cases follow from the definition of liftings of fibred adjunction models.
	\qed
\end{proof}

Before proceeding to the soundness of the refinement type system, we add another assumption that the judgement for logical implication $\Gamma; v : A_u \mid p \vdash q$ implies $\pullback{\pi_{\llbracket \ElimRefinement{\Gamma}; A_u \rrbracket}} \llbracket \Gamma \rrbracket \land \llbracket \ElimRefinement{\Gamma}, v : A_u \vdash p \rrbracket \le \llbracket \ElimRefinement{\Gamma}, v : A_u \vdash q \rrbracket$ in $\category{P}_{\llbracket \ElimRefinement{\Gamma}, v : A_u \rrbracket}$.
One way to achieve this is to collect predicates in $\Gamma$ by
\begin{align*}
	\llparenthesis \emptyctx \rrparenthesis &= \top &
	\llparenthesis \Gamma, x : A \rrparenthesis &= \begin{cases}
		\llparenthesis \Gamma \rrparenthesis \land p[x/v] & \text{if $A = \{ v : A_u \mid p \}$} \\
		\llparenthesis \Gamma \rrparenthesis & \text{otherwise}
	\end{cases}
\end{align*}
and check whether $\llbracket \ElimRefinement{\Gamma}, v : A_u \vdash \llparenthesis \Gamma \rrparenthesis \land p \rrbracket \le \llbracket \ElimRefinement{\Gamma}, v : A_u \vdash q \rrbracket$ holds.
This is sound because we can prove $\llbracket \Gamma \rrbracket \le \llbracket \ElimRefinement{\Gamma} \vdash \llparenthesis \Gamma \rrparenthesis \rrbracket$ by induction on the length of $\Gamma$.
For example, \cite{vazou2014} uses similar encoding of contexts into logical formulas.

\begin{mytheorem}[Soundness]\label{thm:soundness}
	Assume that
	$\llbracket b \rrbracket \in \category{E}_{\ComprehensionFunctor{\llbracket \emptyctx; A \rrbracket}}$ holds for each $b : A \to \mathrm{Type}$ if $\llbracket \emptyctx; A \rrbracket$ is defined, and
	$\llbracket c \rrbracket \in \RefinedCCompC{\category{E}}{\category{P}}_1(1, \llbracket \emptyctx; \mathrm{ty}(c) \rrbracket)$ holds if $\llbracket \emptyctx; \mathrm{ty}(c) \rrbracket \in \RefinedCCompC{\category{E}}{\category{P}}_1$ is defined.
	Then we have the following.
	\begin{itemize}
		\item If $\WellFormedContext{\Gamma}$, then $\llbracket \Gamma \rrbracket \in \category{P}$ is defined.
			If $\WellFormedType{\Gamma}{A}$, then $\llbracket \Gamma; A \rrbracket \in \RefinedCCompC{\category{E}}{\category{P}}_{\llbracket \Gamma \rrbracket}$ is defined.
			If $\WellFormedType{\Gamma}{\underline{C}}$, then $\llbracket \Gamma; \underline{C} \rrbracket \in \category{D}_{\llbracket \Gamma \rrbracket}$ is defined.
		\item If $\ContextSubtyping{\Gamma_1}{\Gamma_2}$, then $\llbracket \Gamma_1 \rrbracket \le \llbracket \Gamma_2 \rrbracket$ in a fibre category of $\category{P}$.
		\item If $\Subtyping{\Gamma}{A}{B}$, then $\llbracket \Gamma; A \rrbracket \subtype \llbracket \Gamma; B \rrbracket$.
			If $\Subtyping{\Gamma}{\underline{C}}{\underline{D}}$, then $\dot{U} \llbracket \Gamma; \underline{C} \rrbracket \subtype \dot{U} \llbracket \Gamma; \underline{D} \rrbracket$.
		\item If $\WellTypedTerm{\Gamma}{V}{A}$, then there exists a lifting $\llbracket \Gamma; V \rrbracket : 1 \llbracket \Gamma \rrbracket \to \llbracket \Gamma; A \rrbracket$ above $\llbracket \ElimRefinement{\Gamma}; V \rrbracket$ along $u : \RefinedCCompC{\category{E}}{\category{P}} \to \category{E}$.
			If $\WellTypedTerm{\Gamma}{M}{\underline{C}}$, then there exists a lifting $\llbracket \Gamma; M \rrbracket : 1 \llbracket \Gamma \rrbracket \to \llbracket \Gamma; \underline{C} \rrbracket$ above $\llbracket \ElimRefinement{\Gamma}; M \rrbracket$ along $u : \RefinedCCompC{\category{E}}{\category{P}} \to \category{E}$.
	\end{itemize}
\end{mytheorem}

Since we have the bijection $s : \RefinedCCompC{\category{E}}{\category{P}}_P(1 P, (X, P, Q)) \to \{ f : P \to Q \mid \pi_{(X, P, Q)} \comp f = \identity{P} \}$ for each $(X, P, Q) \in \RefinedCCompC{\category{E}}{\category{P}}$, we obtain liftings of interpretations of terms along $q : \category{P} \to \category{B}$.
\begin{mycorollary}
	If $\WellTypedTerm{\Gamma}{V}{A}$, then $s \llbracket \ElimRefinement{\Gamma}; V \rrbracket : \llbracket \ElimRefinement{\Gamma} \rrbracket \to \ComprehensionFunctor{\llbracket \ElimRefinement{\Gamma}; A \rrbracket}$ has a lifting $s \llbracket \Gamma; V \rrbracket : \llbracket \Gamma \rrbracket \to \ComprehensionFunctor{\llbracket \Gamma; A \rrbracket}$ along $q : \category{P} \to \category{B}$ (and similarly for computation terms $\WellTypedTerm{\Gamma}{M}{\underline{C}}$).
	\qed
\end{mycorollary}
\begin{mycorollary}
	Assume the lifting of fibred adjunction models is given by applying the Eilenberg-Moore construction to a lifting of monads in Theorem~\ref{thm:refined_monad}.
	If $\WellTypedTerm{\Gamma}{M}{FA}$, then $\theta \comp s \llbracket \ElimRefinement{\Gamma}; M \rrbracket : \llbracket \ElimRefinement{\Gamma} \rrbracket \to T \ComprehensionFunctor{\llbracket \ElimRefinement{\Gamma}; A \rrbracket}$ has a lifting of type $\llbracket \Gamma \rrbracket \to \dot{T} \ComprehensionFunctor{\llbracket \Gamma; A \rrbracket}$ along $q : \category{P} \to \category{B}$.
	\qed
\end{mycorollary}

\section{Toward Recursion in Refinement Type Systems}\label{sec:recursion}
We consider how to deal with general recursion in refinement type systems.
In~\cite{ahman2016}, Ahman used a specific model of the fibration $\mathbf{CFam}(\CPO) \to \CPO$ of continuous families of $\omega$-cpos to extend EMLTT with recursion.
However, we need to identify the structure that characterizes recursion to lift recursion from the underlying type system to refinement type systems.
So, we consider a generalization of Conway operators~\cite{simpson2000} and prove the soundness of the underlying and the refinement type system extended with typing rules for recursion.
This extension enables us to reason about partial correctness of general recursion.

Unfortunately, we still do not know an example of liftings of Conway operators, although (1) $\mathbf{CFam}(\CPO) \to \CPO$ does have a Conway operator and (2) the soundness of the refinement type system with recursion holds under the existence of a lifting of Conway operators.
We leave this problem for future work.

\subsection{Conway Operators}
%First, we recall the notion of Conway operators for Cartesian categories.
%\begin{mydefinition}[Conway operator for Cartesian categories~\cite{simpson2000}]\label{def:conway_cartesian}
%	Let $\category{B}$ be a Cartesian category.
%	A \emph{Conway operator} is a family of mappings $({-})^{\dagger} : \category{B}(I \times X, X) \to \category{B}(I, X)$ satisfying the following.
%	\begin{description}
%		\item[(Naturality)] For each $u : I \to J$ and $f : J \times X \to X$, $f^{\dagger} \comp u = (f \comp (u \times \identity{X}))^{\dagger}$.
%		\item[(Parameterized dinaturality)] For each $f : I \times X \to Y$ and $g : I \times Y \to X$, $(g \comp \langle \pi_1, f \rangle) = g \comp \langle \identity{I}, (f \comp \langle \pi_1, g \rangle)^{\dagger} \rangle$.
%		\item[(Diagonal property)] For each $f : I \times X \times X \to X$, $(f^{\dagger})^{\dagger} = (f \comp (\identity{I} \times \Delta))^{\dagger}$ where $\Delta : X \to X \times X$ is the diagonal map.
%	\end{description}
%\end{mydefinition}
The notion of Conway operators for cartesian categories is defined in~\cite{simpson2000}.
We adapt the definition for comprehension categories with unit.
We allow partially defined Conway operators because we need those defined only on interpretations of computation types.
\begin{mydefinition}[Conway operator for comprehension categories with unit]\label{def:conway_ccompc}
	Let $p : \category{E} \to \category{B}$ be a comprehension category with unit and $K \subseteq \category{E}$ be a collection of objects.
	A \emph{Conway operator} for the comprehension category with unit $p$ defined on $K$ is a family of mappings $({-})^{\ddagger} : \category{E}_I(X, X) \to \category{E}_I(1 I, X)$ for each $X \in \category{E}_I \cap K$ such that the following conditions are satisfied.
	\begin{description}
		\item[(Naturality)] For each $X \in K$, $f \in \category{E}_I(X, X)$, and $u : J \to I$, $\pullback{u} f^{\ddagger} = (\pullback{u} f)^{\ddagger}$.
		\item[(Dinaturality)] For each $X, Y \in K$, $f \in \category{E}_I(X, Y)$, and $g \in \category{E}_I(Y, X)$, $(g \comp f)^{\ddagger} = g \comp (f \comp g)^{\ddagger}$.
		\item[(Diagonal property)] For each $X \in K$ and $f \in \category{E}_{\ComprehensionFunctor{X}}(\pullback{\pi_X} X, \pullback{\pi_X} X)$, if $\pullback{\pi_X} X \in K$, then $(\phi(f^{\ddagger}))^\ddagger = (\phi(\pullback{\delta_X} (\phi^{-1}(f))))^{\ddagger}$ holds where $\phi : \category{E}_{\ComprehensionFunctor{X}}(1 \ComprehensionFunctor{X}, \pullback{\pi_X} X) \to \category{E}_{I}(X, X)$ is the isomorphism defined in \S\ref{sec:preliminaries}.
	\end{description}
\end{mydefinition}

\begin{mylemma}
	Let $\category{B}$ be a cartesian category.
	There is a bijective correspondence between the following.
	(1) Conway operators $({-})^{\dagger}$ on the cartesian category $\category{B}$.
	(2) Conway operators $({-})^{\ddagger}$ on the simple comprehension category $\mathbf{s}(\category{B}) \to \category{B}^{\to}$ that are defined totally on $\mathbf{s}(\category{B})$.
	\qed
\end{mylemma}

\begin{myexample}
	Let $K \subseteq \mathbf{CFam}(\CPO)$ be a collection of objects defined by $K = \{ (I, X) \in \mathbf{CFam}(\CPO) \mid \text{for each $i \in I$, $X i$ has a least element} \}$.
	For each $(I, X) \in K$ and vertical morphism $f = (\identity{I}, (f_i)_{i \in I}) : (I, X) \to (I, X)$, we define $f^{\ddagger} = (\identity{I}, (* \mapsto \mathrm{lfp} f_i)_{i \in I}) : (I, 1) \to (I, X)$.
	Then $({-})^{\ddagger}$ is a Conway operator, which is implicitly used in~\cite{ahman2016}.
 \end{myexample}

\subsection{Recursion in the Underlying Type System}
\paragraph{Syntax.}
We add recursion $\mu x : U \underline{C}. M$ to the syntax of computation terms.
We also add typing rules in Fig.~\ref{fig:typing_recursion}.
\begin{figure}[tb]
	\begin{mathpar}
		\inferrule{
			\WellFormedType{\Gamma}{\underline{C}} \\
			\WellTypedTerm{\Gamma, x : U \underline{C}}{M}{\underline{C}}
		}{\WellTypedTerm{\Gamma}{\mu x : U \underline{C}. M}{\underline{C}}}
		\and
		\inferrule[]{
			\TypeEqual{\Gamma}{\underline{C}}{\underline{D}} \\
			\TermEqual{\Gamma, x : U \underline{C}}{M}{N}{\underline{C}}
		}{\TermEqual{\Gamma}{\mu x : U \underline{C}. M}{\mu x : U \underline{D}. N}{\underline{C}}}
		\and
		\inferrule{
			\WellFormedType{\Gamma}{\underline{C}} \\
			\WellTypedTerm{\Gamma, x : U \underline{C}}{M}{\underline{C}}
		}{\begin{split}
			\TermEqual{\Gamma}{&M[\thunk{(\mu x : U \underline{C}. M)}/x]\\&\qquad}{\mu x : U \underline{C}. M}{\underline{C}}
		\end{split}}
		\and
		\inferrule{
			\WellFormedType{\Gamma}{\underline{C}} \\
			\WellTypedTerm{\Gamma, x : U \underline{C}, y : U \underline{C}}{M}{\underline{C}}
		}{
			\begin{split}
				\TermEqual{\Gamma}{&\mu x : U \underline{C}. \mu y : U \underline{C}. M\\&\qquad}{\mu x : U \underline{C}. M[x/y]}{\underline{C}}
			\end{split}
		}
	\end{mathpar}
	\caption{Typing rules for general recursion.}
	\label{fig:typing_recursion}
\end{figure}

\paragraph{Semantics.}
Assume we have a fibred adjunction model $F \dashv U : r \to p$ where $p : \category{E} \to \category{B}$ and $r : \category{C} \to \category{B}$.
We need a Conway operator defined on objects in $\{ \llbracket \Gamma; U \underline{C} \rrbracket \mid \WellFormedType{\Gamma}{\underline{C}} \} \subseteq \category{E}$.
However, here is a circular definition because $\llbracket \Gamma; U \underline{C} \rrbracket$ may contain terms of the form $\mu x : U \underline{D}. M$, whose interpretations are defined by the Conway operator.
So, we use a slightly stronger condition.
\begin{mydefinition}
	A \emph{Conway operator defined on computation types} is a Conway operator defined on $K \subseteq \category{E}$ such that $K$ satisfies the following conditions.
	(1) $U F X \in K$ holds for each $X \in \category{E}$.
	(2) $\prod_X Y \in K$ holds for each $X \in \category{E}$ and $Y \in K \cap \category{E}_{\ComprehensionFunctor{X}}$.
	(3) For each $X \in K$ and $Y \in \category{E}$, $X \cong Y$ implies $Y \in K$.
\end{mydefinition}

Given a Conway operator defined on computation types, we interpret $\mu x : U \underline{C}. M$ by
$\llbracket \Gamma; \mu x : U \underline{C}. M \rrbracket = \left(\phi(\llbracket \Gamma, x : U \underline{C}; M \rrbracket)\right)^{\ddagger} : 1 \llbracket \Gamma \rrbracket \to U \llbracket \Gamma; \underline{C} \rrbracket$.

\begin{myproposition}
	Soundness (Proposition~\ref{prop:soundness_underlying}) holds for the underlying type system extended with general recursion.
\end{myproposition}
\begin{proof}
	By induction.
	We can prove that the given Conway operator is defined on $\{ \llbracket \Gamma; U \underline{C} \rrbracket \mid \WellFormedType{\Gamma}{\underline{C}} \} \subseteq \category{E}$ by~\cite[Proposition~4.1.14]{ahman2017}.
	\qed
\end{proof}

\subsection{Recursion in Refinement Type System}
\paragraph{Syntax.}
We add the typing rule for $\WellTypedTerm{\Gamma}{\mu x {:} U \underline{C}. M}{\underline{C}}$ in Fig.~\ref{fig:typing_recursion} to the refinement type system.
Here, recall that we remove definitional equalities when we consider the refinement type system.

\paragraph{Semantics.}
We consider liftings of Conway operators to interpret recursion in the refinement type system.
\begin{mydefinition}
	Let $p : \category{E} \to \category{B}$ and $q : \category{D} \to \category{A}$ be comprehension categories with unit, $(u, v) : p \to q$ be a morphism of comprehension categories with unit.
	Assume $q$ has a Conway operator $({-})^{\ddagger}$ defined on $K \subseteq \category{D}$.
	A \emph{lifting} of the Conway operator $({-})^{\ddagger}$ along $(u, v)$ is a Conway operator $({-})^{\natural}$ for $p$ defined on $L \subseteq \category{E}$ such that $u L \subseteq K$ and $u (f^{\natural}) = (u f)^{\ddagger}$ for each $f \in \category{E}_I(X, X)$ where $X \in L$.
\end{mydefinition}
\begin{mylemma}
	Let $(u, v)$ be a morphism of CCompCs defined in Theorem~\ref{thm:ccompc4refinement}.
	Assume $p : \category{E} \to \category{B}$ has a Conway operator $({-})^{\ddagger}$ defined on $K \subseteq \category{E}$.
	The CCompC $\RefinedCCompC{\category{E}}{\category{P}} \to \category{P}$ has a lifting of the Conway operator defined on $L \subseteq \RefinedCCompC{\category{E}}{\category{P}}$ if $u L \subseteq K$ and for each $(X, P, Q) \in L$ and $f \in \RefinedCCompC{\category{E}}{\category{P}}_P((X, P, Q), (X, P, Q))$, $\ComprehensionFunctor{f^{\ddagger}}$ has a lifting $\pullback{\pi_{1 p X}} P \to Q$ along $q : \category{P} \to \category{B}$.
	\qed
\end{mylemma}
\begin{proof}
	Let $(f, \identity{P}, h) : (X, P, Q) \to (X, P, Q)$ be a morphism in $\RefinedCCompC{\category{E}}{\category{P}}$ where $(X, P, Q) \in L$.
	We define a Conway operator by $(f, \identity{P}, h)^{\natural} = (f^{\ddagger}, \identity{P}, h') : (1 p X, P, \pullback{\pi_{1 p X}} P) \to (X, P, Q)$ where $h'$ is a lifting of $\ComprehensionFunctor{f^{\ddagger}}$.
	\qed
\end{proof}

We assume that a lifting of fibred adjunction models~\eqref{eq:lifting_fibred_adjunction_model} together with a lifting of Conway operators defined on computation types is given.
\begin{mytheorem}
	Soundness (Theorem~\ref{thm:soundness}) holds for the refinement type system extended with general recursion.
	\qed
\end{mytheorem}

Consider the fibration $\mathbf{CFam}(\CPO) \to \CPO$ for the underlying type system with recursion.
To support recursion in our refinement type system, a natural choice of a fibration for predicate logic is the fibration of admissible subsets $\mathbf{Adm}(\CPO) \to \CPO$ because the least fixed point of an $\omega$-continuous function $f : X \to X$ is given by $\mathrm{lfp} f = \bigvee_n f^n(\bot)$.
However, we cannot apply Theorem~\ref{thm:ccompc4refinement} because $\mathbf{Adm}(\CPO) \to \CPO$ is not a fibred ccc~\cite[\S{}4.3.2]{hermida1993}.
Specifically, it is not clear whether this combination admits products.
We believe that our approach is quite natural but leave giving concrete examples of liftings of Conway operators for future work.

\section{Related Work}\label{sec:related}
\paragraph{Dependent refinement types.}
Historically, there are two kinds of refinement types.
One is \emph{datasort refinement types}~\cite{freeman1991}, which are subsets of underlying types but not necessarily dependent.
The other is \emph{index refinement types}~\cite{xi1998}.
A typical example of index refinement types is a type of lists indexed by natural numbers that represent the length of lists.
Nowadays, the word ``refinement types'' includes datasort and index refinement types, and moreover, mixtures of them.

Among a wide variety of the meaning of refinement types, we focus on types equipped with predicates that may depend on other terms~\cite{rushby1998,flanagan2006}, which we call \emph{dependent refinement types} or just \emph{refinement types}.
Dependent refinement types are widely studied~\cite{knowles2008,unno2018,barthe2015,lehmann2017}, and implemented in, e.g., F${}^\star$~\cite{swamy2013,swamy2013a} and LiquidHaskell~\cite{rondon2008,vazou2013,vazou2014}.
However, most studies focus on decidable type systems, and only a few consider categorical semantics.

We expect that some of the existing refinement type systems are combined with effect systems.
For example, a dependent refinement type system for nondeterminism and partial/total correctness proposed in~\cite{unno2018} contains types for computations indexed by quantifiers $Q_1 Q_2$ where $Q_1, Q_2 \in \{ \forall, \exists \}$.
Here, $Q_1$ represents may/must nondeterminism, and $Q_2$ represents total/partial correctness.
It has been shown that $Q_1 Q_2$ corresponds to four cartesian liftings of the monad $P_{+}(({-}) + 1)$~\cite{aguirre2020,katsumata_private}.
We conjecture that these liftings are connected by monad morphisms and hence yield a lattice-graded monad.
Another example is a relational refinement type system for differential privacy~\cite{barthe2015}.
Their system seems to use a graded lifting of the distribution monad where the lifting is graded by privacy parameters, as pointed out in~\cite{sato2019}.
We leave for future work combining our refinement type system with effect systems based on graded monads~\cite{katsumata2014,mcdermott2019,fujii2016}.

\paragraph{Categorical semantics.}
Our interpretation of refinement type systems is based on a morphism of CCompCs, which is a similar strategy to~\cite{mellies2015}.
The difference is that our paper focuses on dependent refinement types and makes the role of predicate logic explicit by giving a semantic construction of refinement type systems from given underlying type systems and predicate logic.

Combining dependent types and computational effects is discussed in~\cite{ahman2016,ahman2017,ahman2018}.
Although their aim is not at refinement types, their system is a basis for the design and semantics of our refinement type system with computational effects.

Semantics for types of the form $\{ v : A_u \mid p \}$ are characterized categorically as right adjoints of terminal object functors in~\cite[Chapter~11]{jacobs2001}.
Such types are called \emph{subset types} there.
They consider the situation where a given CCompC $p : \category{E} \to \category{B}$ is already rich enough to interpret $\{ v : A_u \mid p \}$, and do not aim to interpret refinement type systems by liftings of CCompCs.
Moreover, we cannot directly use the interpretations in~\cite{jacobs2001} for our CCompC $\RefinedCCompC{\category{E}}{\category{P}} \to \category{P}$ because we are not given a fibration for predicate logic whose base category is $\category{P}$.
%The problem whether we can define interpretations of $\{ v : A_u \mid p \}$ in this way is left for future work.

%\cite{aguirre2020}

\section{Conclusion and Future Work}\label{sec:conclustion_future_work}
% conclusion
We provided a general construction of liftings of CCompCs from combinations of CCompCs and posetal fibrations satisfying certain conditions.
This can be seen as a semantic counterpart of constructing dependent refinement type systems from underlying type systems and predicate logic.
We identified sufficient conditions for several structures in underlying type systems (e.g.\ products, coproducts, fibred coproducts, fibred monads, and Conway operators) to lift to refinement type systems.
We proved the soundness of a dependent refinement type system with computational effects with respect to interpretations in CCompCs obtained from the general construction.

% future work
We aim to extend our dependent refinement type system by combining effect systems based on graded monads~\cite{katsumata2014,mcdermott2019,fujii2016}.
We hope that this extension will give us a more expressive framework that subsumes, for example, refinement type systems in~\cite{unno2018,barthe2015}.
Another direction is to define interpretations of $\{ v : A_u \mid p \}$ in the style of subset types in~\cite[Chapter~11]{jacobs2001}.
Lastly, we are interested in finding more examples of possible combinations of underlying type systems and predicate logic (especially for recursion in refinement type systems but not limited to this) so that we can find a new practical application of this paper.

\bibliographystyle{splncs04}
\bibliography{ref}

\begin{thebibliography}{10}
\providecommand{\url}[1]{\texttt{#1}}
\providecommand{\urlprefix}{URL }
\providecommand{\doi}[1]{https://doi.org/#1}

\bibitem{aguirre2020}
Aguirre, Katsumata, S.: Weakest preconditions in fibrations. In: Proceedings of
  the Thirty-Sixth Conference on the Mathematical Foundations of Programming
  Semantics, {MFPS} 2020, Paris, France (June 2020), to appear

\bibitem{ahman2017}
Ahman, D.: Fibred {{Computational Effects}}. {{PhD Thesis}}, University of
  Edinburgh (2017)

\bibitem{ahman2018}
Ahman, D.: Handling fibred algebraic effects. Proceedings of the ACM on
  Programming Languages  \textbf{2},  1--29 (Jan 2018). \doi{10.1145/3158095}

\bibitem{ahman2016}
Ahman, D., Ghani, N., Plotkin, G.D.: Dependent types and fibred computational
  effects. In: Jacobs, B., Löding, C. (eds.) Foundations of Software Science
  and Computation Structures, vol.~9634, pp. 36--54. {Springer Berlin
  Heidelberg} (2016). \doi{10.1007/978-3-662-49630-5\_3}

\bibitem{barthe2015}
Barthe, G., Gaboardi, M., Gallego~Arias, E.J., Hsu, J., Roth, A., Strub, P.Y.:
  Higher-{{Order Approximate Relational Refinement Types}} for {{Mechanism
  Design}} and {{Differential Privacy}}. In: Proceedings of the 42nd {{Annual
  ACM SIGPLAN}}-{{SIGACT Symposium}} on {{Principles}} of {{Programming
  Languages}} - {{POPL}} '15. pp. 55--68. {ACM Press}, {Mumbai, India} (2015).
  \doi{10.1145/2676726.2677000}

\bibitem{flanagan2006}
Flanagan, C.: Hybrid type checking. In: Conference Record of the 33rd {{ACM
  SIGPLAN}}-{{SIGACT}} Symposium on {{Principles}} of Programming Languages -
  {{POPL}}'06. pp. 245--256. {ACM Press}, {Charleston, South Carolina, USA}
  (2006). \doi{10.1145/1111037.1111059}

\bibitem{freeman1991}
Freeman, T., Pfenning, F.: Refinement types for {{ML}}. ACM SIGPLAN Notices
  \textbf{26}(6),  268--277 (Jun 1991). \doi{10.1145/113446.113468}

\bibitem{fujii2016}
Fujii, S., Katsumata, S.y., Melli{\`e}s, P.A.: Towards a {{Formal Theory}} of
  {{Graded Monads}}. In: Jacobs, B., L{\"o}ding, C. (eds.) Foundations of
  {{Software Science}} and {{Computation Structures}}, vol.~9634, pp. 513--530.
  {Springer Berlin Heidelberg}, {Berlin, Heidelberg} (2016).
  \doi{10.1007/978-3-662-49630-5\_30}

\bibitem{hermida1993}
Hermida, C.: Fibrations, logical predicates and indeterminates. {PhD} {Thesis},
  University of Edinburgh, UK (1993)

\bibitem{jacobs2001}
Jacobs, B.: Categorical Logic and Type Theory. No.~141 in Studies in Logic and
  the Foundations of Mathematics, {Elsevier}, paperback edn. (2001)

\bibitem{katsumata2014}
Katsumata, S.: Parametric effect monads and semantics of effect systems. In:
  Proceedings of the 41st {{ACM SIGPLAN}}-{{SIGACT Symposium}} on
  {{Principles}} of {{Programming Languages}} - {{POPL}} '14. pp. 633--645.
  {ACM Press}, {San Diego, California, USA} (2014).
  \doi{10.1145/2535838.2535846}

\bibitem{katsumata_private}
Katsumata, S.: private communication (2020)

\bibitem{knowles2008}
Knowles, K., Flanagan, C.: Compositional reasoning and decidable checking for
  dependent contract types. In: Proceedings of the 3rd Workshop on
  {{Programming}} Languages Meets Program Verification - {{PLPV}} '09. p.~27.
  {ACM Press}, {Savannah, GA, USA} (2008). \doi{10.1145/1481848.1481853}

\bibitem{lehmann2017}
Lehmann, N., Tanter, {\'E}.: Gradual refinement types. ACM SIGPLAN Notices
  \textbf{52}(1),  775--788 (May 2017). \doi{10.1145/3093333.3009856}

\bibitem{mcdermott2019}
McDermott, D., Mycroft, A.: Extended {{Call}}-by-{{Push}}-{{Value}}:
  {{Reasoning About Effectful Programs}} and {{Evaluation Order}}. In: Caires,
  L. (ed.) Programming {{Languages}} and {{Systems}}, vol. 11423, pp. 235--262.
  {Springer International Publishing}, {Cham} (2019).
  \doi{10.1007/978-3-030-17184-1\_9}

\bibitem{mellies2015}
Melli{\`e}s, P.A., Zeilberger, N.: Functors are {{Type Refinement Systems}}.
  In: Proceedings of the 42nd {{Annual ACM SIGPLAN}}-{{SIGACT Symposium}} on
  {{Principles}} of {{Programming Languages}} - {{POPL}} '15. pp. 3--16. {ACM
  Press}, {Mumbai, India} (2015). \doi{10.1145/2676726.2676970}

\bibitem{moggi1991}
Moggi, E.: Notions of computation and monads. Information and Computation
  \textbf{93}(1),  55--92 (Jul 1991). \doi{10.1016/0890-5401(91)90052-4}

\bibitem{ou2004}
Ou, X., Tan, G., Mandelbaum, Y., Walker, D.: Dynamic {{Typing}} with
  {{Dependent Types}}. In: Levy, J.J., Mayr, E.W., Mitchell, J.C. (eds.)
  Exploring {{New Frontiers}} of {{Theoretical Informatics}}, vol.~155, pp.
  437--450. {Kluwer Academic Publishers}, {Boston} (2004).
  \doi{10.1007/1-4020-8141-3\_34}

\bibitem{rondon2008}
Rondon, P.M., Kawaguci, M., Jhala, R.: Liquid types. In: Proceedings of the
  2008 {{ACM SIGPLAN}} Conference on {{Programming}} Language Design and
  Implementation - {{PLDI}} '08. p.~159. {ACM Press}, {Tucson, AZ, USA} (2008).
  \doi{10.1145/1375581.1375602}

\bibitem{rushby1998}
Rushby, J., Owre, S., Shankar, N.: Subtypes for specifications: Predicate
  subtyping in {{PVS}}. IEEE Transactions on Software Engineering
  \textbf{24}(9),  709--720 (Sept/1998). \doi{10.1109/32.713327}

\bibitem{sato2019}
Sato, T., Barthe, G., Gaboardi, M., Hsu, J., Katsumata, S.y.: Approximate
  {{Span Liftings}}: {{Compositional Semantics}} for {{Relaxations}} of
  {{Differential Privacy}}. In: 2019 34th {{Annual ACM}}/{{IEEE Symposium}} on
  {{Logic}} in {{Computer Science}} ({{LICS}}). pp. 1--14. {IEEE}, {Vancouver,
  BC, Canada} (Jun 2019). \doi{10.1109/LICS.2019.8785668}

\bibitem{simpson2000}
Simpson, A., Plotkin, G.: Complete axioms for categorical fixed-point
  operators. In: Proceedings {{Fifteenth Annual IEEE Symposium}} on {{Logic}}
  in {{Computer Science}} ({{Cat}}. {{No}}.{{99CB36332}}). pp. 30--41. {IEEE
  Comput. Soc}, {Santa Barbara, CA, USA} (2000). \doi{10.1109/LICS.2000.855753}

\bibitem{swamy2013}
Swamy, N., Chen, J., Fournet, C., Strub, P.Y., Bhargavan, K., Yang, J.: Secure
  distributed programming with value-dependent types. Journal of Functional
  Programming  \textbf{23}(4),  402--451 (Jul 2013).
  \doi{10.1017/S0956796813000142}

\bibitem{swamy2013a}
Swamy, N., Weinberger, J., Schlesinger, C., Chen, J., Livshits, B.: Verifying
  higher-order programs with the dijkstra monad. In: Proceedings of the 34th
  {{ACM SIGPLAN}} Conference on {{Programming}} Language Design and
  Implementation - {{PLDI}} '13. p.~387. {ACM Press}, {Seattle, Washington,
  USA} (2013). \doi{10.1145/2491956.2491978}

\bibitem{unno2018}
Unno, H., Satake, Y., Terauchi, T.: Relatively complete refinement type system
  for verification of higher-order non-deterministic programs. Proceedings of
  the ACM on Programming Languages  \textbf{2},  1--29 (Jan 2018).
  \doi{10.1145/3158100}

\bibitem{vazou2013}
Vazou, N., Rondon, P.M., Jhala, R.: Abstract {{Refinement Types}}. In:
  Hutchison, D., Kanade, T., Kittler, J., Kleinberg, J.M., Mattern, F.,
  Mitchell, J.C., Naor, M., Nierstrasz, O., Pandu~Rangan, C., Steffen, B.,
  Sudan, M., Terzopoulos, D., Tygar, D., Vardi, M.Y., Weikum, G., Felleisen,
  M., Gardner, P. (eds.) Programming {{Languages}} and {{Systems}}, vol.~7792,
  pp. 209--228. {Springer Berlin Heidelberg}, {Berlin, Heidelberg} (2013).
  \doi{10.1007/978-3-642-37036-6\_13}

\bibitem{vazou2014}
Vazou, N., Seidel, E.L., Jhala, R., Vytiniotis, D., Peyton-Jones, S.:
  Refinement types for {Haskell}. In: Proceedings of the 19th {ACM} {SIGPLAN}
  international conference on {Functional} programming - {ICFP} '14. pp.
  269--282. ACM Press, Gothenburg, Sweden (2014). \doi{10.1145/2628136.2628161}

\bibitem{xi1998}
Xi, H., Pfenning, F.: Eliminating array bound checking through dependent types.
  In: Proceedings of the {{ACM SIGPLAN}} 1998 Conference on {{Programming}}
  Language Design and Implementation - {{PLDI}} '98. pp. 249--257. {ACM Press},
  {Montreal, Quebec, Canada} (1998). \doi{10.1145/277650.277732}

\end{thebibliography}

\appendix
\section{Underlying Type System}
\subsection{Typing Rules}

\subsubsection{Well-Formed Contexts}

\begin{mathpar}
    \inferrule{ }{\WellFormedContext{\emptyctx}}
    \and
    \inferrule{
        \WellFormedContext{\Gamma} \\
        \WellFormedType{\Gamma}{A} \\
        x \notin \mathrm{Vars}(\Gamma)
    }{\WellFormedContext{\Gamma, x : A}}
\end{mathpar}

\subsubsection{Definitional Equality for Contexts}
\begin{mathpar}
    \inferrule{ }{\ContextEqual{\emptyctx}{\emptyctx}}
    \and
    \inferrule{
        \ContextEqual{\Gamma_1}{\Gamma_2} \\
        \TypeEqual{\Gamma_1}{A}{B} \\
        \WellFormedType{\Gamma_2}{B} \\
        x \notin \mathrm{Vars}(\Gamma_1) \cup \mathrm{Vars}(\Gamma_2)
    }{\ContextEqual{\Gamma_1, x : A}{\Gamma_2, x : B}}
\end{mathpar}

\subsubsection{Definitional Equality}
\paragraph{Reflexivity.}
\begin{mathpar}
    \inferrule{\WellFormedType{\Gamma}{A}}{\TypeEqual{\Gamma}{A}{A}}
    \and
    \inferrule{\WellFormedType{\Gamma}{\underline{C}}}{\TypeEqual{\Gamma}{\underline{C}}{\underline{C}}}
    \and
    \inferrule{\WellTypedTerm{\Gamma}{V}{A}}{\TermEqual{\Gamma}{V}{V}{A}}
    \and
    \inferrule{\WellTypedTerm{\Gamma}{M}{\underline{C}}}{\TermEqual{\Gamma}{M}{M}{\underline{C}}}
\end{mathpar}

\paragraph{Symmetry.}
\begin{mathpar}
    \inferrule{\TypeEqual{\Gamma}{B}{A}}{\TypeEqual{\Gamma}{A}{B}}
    \and
    \inferrule{\TypeEqual{\Gamma}{\underline{D}}{\underline{C}}}{\TypeEqual{\Gamma}{\underline{C}}{\underline{D}}}
    \and
    \inferrule{\TermEqual{\Gamma}{W}{V}{A}}{\TermEqual{\Gamma}{V}{W}{A}}
    \and
    \inferrule{\TermEqual{\Gamma}{N}{M}{\underline{C}}}{\TermEqual{\Gamma}{M}{N}{\underline{C}}}
\end{mathpar}

\paragraph{Transitivity.}
\begin{mathpar}
    \inferrule{
        \TermEqual{\Gamma}{V_1}{V_2}{A} \\
        \TermEqual{\Gamma}{V_2}{V_3}{A}
    }{\TermEqual{\Gamma}{V_1}{V_3}{A}}
    \and
    \inferrule{
        \TermEqual{\Gamma}{M_1}{M_2}{\underline{C}} \\
        \TermEqual{\Gamma}{M_2}{M_3}{\underline{C}}
    }{\TermEqual{\Gamma}{M_1}{M_3}{\underline{C}}}
    \and
    \inferrule{
        \TypeEqual{\Gamma}{A_1}{A_2} \\
        \TypeEqual{\Gamma}{A_2}{A_3}
    }{\TypeEqual{\Gamma}{A_1}{A_3}}
    \and
    \inferrule{
        \TypeEqual{\Gamma}{\underline{C}_1}{\underline{C}_2} \\
        \TypeEqual{\Gamma}{\underline{C}_2}{\underline{C}_3}
    }{\TypeEqual{\Gamma}{\underline{C}_1}{\underline{C}_3}}
\end{mathpar}

\subsubsection{Conversion}
\begin{mathpar}
    \inferrule{
        \WellFormedType{\Gamma_2}{A} \\
        \ContextEqual{\Gamma_1}{\Gamma_2}
    }{\WellFormedType{\Gamma_1}{A}}
    \and
    \inferrule{
        \WellFormedType{\Gamma_2}{\underline{C}} \\
        \ContextEqual{\Gamma_1}{\Gamma_2}
    }{\WellFormedType{\Gamma_1}{\underline{C}}}
    \and
    \inferrule{
        \TypeEqual{\Gamma_2}{A}{B} \\
        \ContextEqual{\Gamma_1}{\Gamma_2}
    }{\TypeEqual{\Gamma_1}{A}{B}}
    \and
    \inferrule{
        \TypeEqual{\Gamma_2}{\underline{C}}{\underline{D}} \\
        \ContextEqual{\Gamma_1}{\Gamma_2}
    }{\TypeEqual{\Gamma_1}{\underline{C}}{\underline{D}}}
    \and
    \inferrule{
        \ContextEqual{\Gamma_1}{\Gamma_2} \\
        \WellTypedTerm{\Gamma_2}{V}{A} \\
        \TypeEqual{\Gamma_1}{A}{B}
    }{\WellTypedTerm{\Gamma_1}{V}{B}}
    \and
    \inferrule{
        \ContextEqual{\Gamma_1}{\Gamma_2} \\
        \WellTypedTerm{\Gamma_2}{M}{\underline{C}} \\
        \TypeEqual{\Gamma_1}{\underline{C}}{\underline{D}}
    }{\WellTypedTerm{\Gamma_1}{M}{\underline{D}}}
\end{mathpar}

\subsubsection{Variables}
\begin{mathpar}
    \inferrule{\WellFormedContext{\Gamma_1, x : A, \Gamma_2}}{\WellTypedTerm{\Gamma_1, x : A, \Gamma_2}{x}{A}}
\end{mathpar}

\subsubsection{Value constants}
\begin{mathpar}
    \inferrule{
        \WellFormedContext{\Gamma} \\
        \WellFormedType{\emptyctx}{\mathrm{ty}(c)}
    }{\WellTypedTerm{\Gamma}{c_{\mathrm{ty}(c)}}{\mathrm{ty}(c)}}
\end{mathpar}

\subsubsection{Unit Type}
\begin{mathpar}
    \inferrule{\WellFormedContext{\Gamma}}{\WellFormedType{\Gamma}{1}}
    \and
    \inferrule{\WellFormedContext{\Gamma}}{\WellTypedTerm{\Gamma}{*}{1}}
    \and
    \inferrule{\WellTypedTerm{\Gamma}{V}{1}}{\TermEqual{\Gamma}{V}{*}{1}}
\end{mathpar}

\subsubsection{Base Types}
\begin{mathpar}
	\inferrule{
		\WellTypedTerm{\Gamma}{V}{A} \\
		b : A \to \mathrm{Type} \\
		\WellFormedType{\emptyctx}{A}
	}{\WellFormedType{\Gamma}{b_A(V)}}
    \and
	\inferrule{
		\TermEqual{\Gamma}{V}{W}{A} \\
		b : A \to \mathrm{Type} \\
		\WellFormedType{\emptyctx}{A}
	}{\TypeEqual{\Gamma}{b_A(V)}{b_A(W)}}
\end{mathpar}

\subsubsection{Value $\Sigma$-Types}
\begin{mathpar}
    \inferrule{
        \WellFormedType{\Gamma}{A} \\
        \WellFormedType{\Gamma, x : A}{B}
    }{\WellFormedType{\Gamma}{\ValueSumType{x}{A}{B}}}
    \and
    \inferrule{
        \WellTypedTerm{\Gamma}{V}{A} \\
        \WellFormedType{\Gamma, x : A}{B} \\
        \WellTypedTerm{\Gamma}{W}{B[V/x]}
    }{\WellTypedTerm{\Gamma}{\langle V, W \rangle}{\ValueSumType{x}{A}{B}}}
    \and
    \inferrule{
        \WellTypedTerm{\Gamma}{V}{\ValueSumType{x}{A}{B}} \\
        \WellFormedType{\Gamma, z : \ValueSumType{x}{A}{B}}{\underline{C}} \\
        \WellTypedTerm{\Gamma, x : A, y : B}{M}{\underline{C}[\langle x, y \rangle / z]}
    }{\WellTypedTerm{\Gamma}{\PatternMatch{V}{x}{A}{y}{B}{z}{\underline{C}}{M}}{\underline{C}[V/z]}}
\end{mathpar}
\begin{mathpar}
    \inferrule{
        \TypeEqual{\Gamma}{A_1}{A_2} \\
        \TypeEqual{\Gamma, x : A_1}{B_1}{B_2}
    }{\TypeEqual{\Gamma}{\Sigma x : A_1. B_1}{\Sigma x : A_2. B_2}}
    \and
	\inferrule{
		\TypeEqual{\Gamma}{A_1}{A_2} \\
		\TermEqual{\Gamma}{V_1}{V_2}{A_2} \\\\
		\TypeEqual{\Gamma, x : A_1}{B_1}{B_2} \\
		\TermEqual{\Gamma}{W_1}{W_2}{B_2[V_2/x]}
	}{\TermEqual{\Gamma}{\langle V_1, W_1 \rangle}{\langle V_2, W_2 \rangle}{\Sigma x : A_2. B_2}}
\end{mathpar}
\begin{mathpar}
	\inferrule{
		\TypeEqual{\Gamma}{A_1}{A_2} \\
		\TypeEqual{\Gamma, x : A_1}{B_1}{B_2} \\
		\TypeEqual{\Gamma, z : \ValueSumType{x}{A_1}{B_1}}{\underline{C}_1}{\underline{C}_2} \\\\
		\TermEqual{\Gamma}{V_1}{V_2}{\ValueSumType{x}{A_2}{B_2}} \\
		\TermEqual{\Gamma, x : A_1, y : B_1}{M_1}{M_2}{\underline{C}_2[\DTuple{x}{y}{x}{A_2}{B_2}/z]}
    }{\TermEqual{\Gamma}{\PatternMatch{V_1}{x}{A_1}{y}{B_1}{z}{\underline{C}_1}{M_1}}{\PatternMatch{V_2}{x}{A_2}{y}{B_2}{z}{\underline{C}_2}{M_2}}{\underline{C}_2[V_2/z]}}
    \and
    \inferrule{
        \WellFormedType{\Gamma, z : \Sigma x {:} A. B}{\underline{C}} \\
        \WellTypedTerm{\Gamma}{V}{A} \\
        \WellTypedTerm{\Gamma}{W}{B[V/x]} \\
        \WellTypedTerm{\Gamma, x : A, y : B}{M}{\underline{C}[\langle x, y \rangle / z]}
    }{\TermEqual{\Gamma}{\PatternMatch{\langle V, W \rangle}{x}{A}{y}{B}{z}{\underline{C}}{M}}{M[V/x][W/y]}{\underline{C}[\langle V, W \rangle / z]}}
    \and
    \inferrule{
		\WellFormedType{\Gamma}{A} \\
		\WellFormedType{\Gamma, x : A}{B} \\
		\WellTypedTerm{\Gamma}{V}{\Sigma x {:} A. B} \\
		\WellFormedType{\Gamma, z : \Sigma x {:} A. B}{\underline{C}} \\
		\WellTypedTerm{\Gamma, z : \Sigma x {:} A. B}{M}{\underline{C}}
	}{\TermEqual{\Gamma}{\PatternMatch{V}{x}{A}{y}{B}{z}{\underline{C}}{M[\langle x, y \rangle / z]}}{M[V/z]}{\underline{C}[V/z]}}
\end{mathpar}

\subsubsection{Thunked Computation}
\begin{mathpar}
    \inferrule{\WellFormedType{\Gamma}{\underline{C}}}{\WellFormedType{\Gamma}{U\underline{C}}}
    \and
    \inferrule{\WellTypedTerm{\Gamma}{M}{\underline{C}}}{\WellTypedTerm{\Gamma}{\mathbf{thunk}\ M}{U \underline{C}}}
    \and
    \inferrule{\WellTypedTerm{\Gamma}{V}{U \underline{C}}}{\WellTypedTerm{\Gamma}{\mathbf{force}_{\underline{C}}\ V}{\underline{C}}}
    \and
    \inferrule{
        \TypeEqual{\Gamma}{\underline{C}_1}{\underline{C}_2}
    }{
        \TypeEqual{\Gamma}{U \underline{C}_1}{U \underline{C}_2}
    }
    \and
    \inferrule{
        \TermEqual{\Gamma}{M_1}{M_2}{\underline{C}}
    }{
        \TermEqual{\Gamma}{\mathbf{thunk}\ M_1}{\mathbf{thunk}\ M_2}{U \underline{C}}
    }
    \and
    \inferrule{
        \TypeEqual{\Gamma}{\underline{C}_1}{\underline{C}_2} \\
        \TermEqual{\Gamma}{V_1}{V_2}{U \underline{C}_2}
    }{
        \TermEqual{\Gamma}{\force{\underline{C}_1}{V_1}}{\force{\underline{C}_2}{V_2}}{\underline{C}_2}
    }
    \and
    \inferrule{
        \WellTypedTerm{\Gamma}{V}{U \underline{C}}
    }{
        \TermEqual{\Gamma}{\thunk{(\force{\underline{C}}{V}})}{V}{U \underline{C}}
    }
    \and
    \inferrule{
        \WellTypedTerm{\Gamma}{M}{\underline{C}}
    }{
        \TermEqual{\Gamma}{\force{\underline{C}}{(\thunk{M}})}{M}{\underline{C}}
    }
\end{mathpar}

\subsubsection{Return}
\begin{mathpar}
    \inferrule{
        \WellFormedType{\Gamma}{A}
    }{
        \WellFormedType{\Gamma}{F A}
    }
    \and
    \inferrule{
        \WellTypedTerm{\Gamma}{V}{A}
    }{
        \WellTypedTerm{\Gamma}{\return{V}}{F A}
    }
    \and
    \inferrule{
        \WellTypedTerm{\Gamma}{M}{F A} \\
        \WellFormedType{\Gamma}{\underline{C}} \\
        \WellTypedTerm{\Gamma, x : A}{N}{\underline{C}}
    }{
        \WellTypedTerm{\Gamma}{\SeqComp{M}{x}{A}{\underline{C}}{N}}{\underline{C}}
    }
\end{mathpar}
\begin{mathpar}
    \inferrule{
        \TypeEqual{\Gamma}{A_1}{A_2}
    }{
        \TypeEqual{\Gamma}{F A_1}{F A_2}
    }
    \and
    \inferrule{
        \TermEqual{\Gamma}{V_1}{V_2}{A}
    }{
        \TermEqual{\Gamma}{\return{V_1}}{\return{V_2}}{F A}
    }
    \and
    \inferrule{
		\TypeEqual{\Gamma}{A_1}{A_2} \\
		\TermEqual{\Gamma}{M_1}{M_2}{F A_2} \\\\
		\TypeEqual{\Gamma}{\underline{C}_1}{\underline{C}_2} \\
		\TermEqual{\Gamma, x : A_1}{N_1}{N_2}{\underline{C}_2}
    }{\TermEqual{\Gamma}{\SeqComp{M_1}{x}{A_1}{\underline{C}_1}{N_1}}{\SeqComp{M_2}{x}{A_2}{\underline{C}_2}{N_2}}{\underline{C_2}}}
\end{mathpar}
\begin{mathpar}
    \inferrule{
        \WellTypedTerm{\Gamma}{V}{A} \\
        \WellFormedType{\Gamma}{\underline{C}} \\
        \WellTypedTerm{\Gamma, x : A}{M}{\underline{C}}
    }{
        \TermEqual{\Gamma}{\SeqComp{\return{V}}{x}{A}{\underline{C}}{M}}{M[V/x]}{\underline{C}[V/x]}
    }
    \and
    \inferrule{
        \WellTypedTerm{\Gamma}{M}{F A}
    }{
        \TermEqual{\Gamma}{\SeqComp{M}{x}{A}{\underline{C}}{\return{x}}}{M}{F A}
    }
    \and
    \inferrule{
        \WellTypedTerm{\Gamma}{M_1}{F A_1} \\
        \WellFormedType{\Gamma}{A_2} \\
        \WellTypedTerm{\Gamma, x : A_1}{M_2}{F A_2} \\
        \WellFormedType{\Gamma}{\underline{C}} \\
        \WellTypedTerm{\Gamma, y : A_2}{M_3}{\underline{C}}
    }{
        \begin{split}
            \TermEqual{\Gamma}{&\SeqComp{(\SeqComp{M_1}{x}{A_1}{F A_2}{M_2})}{y}{A_2}{\underline{C}}{M_3}\\&}{\SeqComp{M_1}{x}{A_1}{F A_2}{(\SeqComp{M_2}{y}{A_2}{\underline{C}}{M_3})}}{\underline{C}}
        \end{split}
    }
\end{mathpar}

\subsubsection{Computational $\Pi$-Types}
\begin{mathpar}
    \inferrule{
        \WellFormedType{\Gamma}{A} \\
        \WellFormedType{\Gamma, x : A}{\underline{C}}
    }{
        \WellFormedType{\Gamma}{\CompProdType{x}{A}{\underline{C}}}
    }
    \and
    \inferrule{
        \WellTypedTerm{\Gamma, x : A}{M}{\underline{C}}
    }{
        \WellTypedTerm{\Gamma}{\lambda x : A. M}{\CompProdType{x}{A}{\underline{C}}}
    }
    \and
    \inferrule{
        \WellFormedType{\Gamma, x : A}{\underline{C}} \\
        \WellTypedTerm{\Gamma}{M}{\CompProdType{x}{A}{\underline{C}}} \\
        \WellTypedTerm{\Gamma}{V}{A}
    }{
        \WellTypedTerm{\Gamma}{M(V)_{(x : A). \underline{C}}}{\underline{C}[V/x]}
    }
\end{mathpar}
\begin{mathpar}
    \inferrule{
        \TypeEqual{\Gamma}{A_2}{A_1} \\
        \TypeEqual{\Gamma, x : A_2}{\underline{C}_1}{\underline{C}_2}
    }{
        \TypeEqual{\Gamma}{\CompProdType{x}{A_1}{\underline{C}_1}}{\CompProdType{x}{A_2}{\underline{C}_2}}
    }
    \and
    \inferrule{
        \TypeEqual{\Gamma}{A_1}{A_2} \\
        \TermEqual{\Gamma, x : A_1}{M_1}{M_2}{\underline{C}}
    }{
        \TermEqual{\Gamma}{\lambda x : A_1. M_1}{\lambda x : A_2. M_2}{\CompProdType{x}{A_1}{\underline{C}}}
    }
    \and
	\inferrule{
		\TypeEqual{\Gamma}{A_2}{A_1} \\
		\TypeEqual{\Gamma, x : A_2}{\underline{C}_1}{\underline{C}_2} \\\\
		\TermEqual{\Gamma}{M_1}{M_2}{\CompProdType{x}{A_2}{\underline{C}_2}} \\
		\TermEqual{\Gamma}{V_1}{V_2}{A_2}
    }{\TermEqual{\Gamma}{M_1(V_1)_{(x : A_1). \underline{C}_1}}{M_2(V_2)_{(x : A_2). \underline{C}_2}}{\underline{C}_1[V_1/x]}}
\end{mathpar}
\begin{mathpar}
    \inferrule{
        \WellTypedTerm{\Gamma, x : A}{M}{\underline{C}} \\
        \WellTypedTerm{\Gamma}{V}{A}
    }{
        \TermEqual{\Gamma}{(\lambda x : A. M)(V)_{(x : A). \underline{C}}}{M[V/x]}{\underline{C}[V/x]}
    }
    \and
    \inferrule{
        \WellFormedType{\Gamma, x : A}{\underline{C}} \\
        \WellTypedTerm{\Gamma}{M}{\CompProdType{x}{A}{\underline{C}}}
    }{
        \TermEqual{\Gamma}{M}{\lambda x : A. M(x)_{(x : A). \underline{C}}}{\CompProdType{x}{A}{\underline{C}}}
    }
\end{mathpar}

\subsubsection{Fibred Coproduct Types}
\begin{mathpar}
    \inferrule{
        \WellFormedType{\Gamma}{A} \\
        \WellFormedType{\Gamma}{B}
    }{
        \WellFormedType{\Gamma}{A + B}
    }
    \and
    \inferrule{
        \WellTypedTerm{\Gamma}{V}{A} \\
        \WellFormedType{\Gamma}{B}
    }{
        \WellTypedTerm{\Gamma}{\mathbf{inl}_{A + B}\ V}{A + B}
    }
    \and
    \inferrule{
        \WellTypedTerm{\Gamma}{V}{B} \\
        \WellFormedType{\Gamma}{A}
    }{
        \WellTypedTerm{\Gamma}{\mathbf{inr}_{A + B}\ V}{A + B}
    }
    \and
    \inferrule{
		\WellFormedType{\Gamma, z : A + B}{\underline{C}} \\
		\WellTypedTerm{\Gamma}{V}{A + B} \\
		\WellTypedTerm{\Gamma, x : A}{M}{\underline{C}[\mathbf{inl}_{A+B}\ x/z]} \\
		\WellTypedTerm{\Gamma, y : B}{N}{\underline{C}[\mathbf{inr}_{A+B}\ y/z]}
    }{\WellTypedTerm{\Gamma}{\mathbf{case}\ V\ \mathbf{of}_{z.\underline{C}}\ (\mathbf{inl}\ (x : A) \mapsto M, \mathbf{inr}\ (y : B) \mapsto N)}{\underline{C}[V/z]}}
\end{mathpar}
\begin{mathpar}
    \inferrule{
        \TypeEqual{\Gamma}{A_1}{A_2} \\
        \TypeEqual{\Gamma}{B_1}{B_2}
    }{
        \TypeEqual{\Gamma}{A_1 + B_1}{A_2 + B_2}
    }
    \and
	\inferrule{
		\TypeEqual{\Gamma}{A_1}{A_2} \\
		\TypeEqual{\Gamma}{B_1}{B_2} \\
		\TermEqual{\Gamma}{V_1}{V_2}{A_2}
	}{\TermEqual{\Gamma}{\mathbf{inl}_{A_1 + B_1}\ V_1}{\mathbf{inl}_{A_2 + B_2}\ V_2}{A_2 + B_2}}
    \and
    \inferrule{
		\TypeEqual{\Gamma}{A_1}{A_2} \\
		\TypeEqual{\Gamma}{B_1}{B_2} \\
		\TermEqual{\Gamma}{V_1}{V_2}{B_2}
	}{\TermEqual{\Gamma}{\mathbf{inr}_{A_1 + B_1}\ V_1}{\mathbf{inr}_{A_2 + B_2}\ V_2}{A_2 + B_2}}
    \and
    \inferrule{
		\TypeEqual{\Gamma}{A_1}{A_2} \\
		\TypeEqual{\Gamma}{B_1}{B_2} \\
		\TypeEqual{\Gamma, z : A_1 + B_1}{\underline{C}_1}{\underline{C}_2} \\
		\TermEqual{\Gamma}{V_1}{V_2}{A_1 + B_1} \\
		\TermEqual{\Gamma, x : A_1}{M_1}{M_2}{\underline{C}_2[\mathbf{inl}_{A_2 + B_2}\ x/z]} \\
		\TermEqual{\Gamma, y : B_1}{N_1}{N_2}{\underline{C}_2[\mathbf{inr}_{A_2 + B_2}\ y/z]}
	}{
        \begin{split}
            \TermEqual{\Gamma}{&\mathbf{case}\ V_1\ \mathbf{of}_{z.\underline{C}_1}\ (\mathbf{inl}\ (x : A_1) \mapsto M_1, \mathbf{inr}\ (y : B_1) \mapsto N_1)\\&}{\mathbf{case}\ V_2\ \mathbf{of}_{z.\underline{C}_2}\ (\mathbf{inl}\ (x : A_2) \mapsto M_2, \mathbf{inr}\ (y : B_2) \mapsto N_2)}{\underline{C}_2[V_2/z]}
        \end{split}
    }
\end{mathpar}
\begin{mathpar}
    \inferrule{
		\WellFormedType{\Gamma, z : A + B}{\underline{C}} \\
		\WellTypedTerm{\Gamma}{V}{A} \\\\
		\WellTypedTerm{\Gamma, x : A}{M}{\underline{C}[\mathbf{inl}_{A+B}\ x/z]} \\
		\WellTypedTerm{\Gamma, y : B}{N}{\underline{C}[\mathbf{inr}_{A+B}\ y/z]}
	}{\TermEqual{\Gamma}{\Case{(\mathbf{inl}_{A+B}\ V)}{z}{\underline{C}}{x}{A}{M}{y}{B}{N}}{M[V/x]}{\underline{C}[\mathbf{inl}_{A+B}\ V/z]}}
    \and
    \inferrule{
		\WellFormedType{\Gamma, z : A + B}{\underline{C}} \\
		\WellTypedTerm{\Gamma}{V}{B} \\\\
		\WellTypedTerm{\Gamma, x : A}{M}{\underline{C}[\mathbf{inl}_{A+B}\ x/z]} \\
		\WellTypedTerm{\Gamma, y : B}{N}{\underline{C}[\mathbf{inr}_{A+B}\ y/z]}
	}{\TermEqual{\Gamma}{\Case{(\mathbf{inr}_{A+B}\ V)}{z}{\underline{C}}{x}{A}{M}{y}{B}{N}}{N[V/y]}{\underline{C}[\mathbf{inr}_{A+B}\ V/z]}}
    \and
    \inferrule{
		\WellFormedType{\Gamma, z : A + B}{\underline{C}} \\
		\WellTypedTerm{\Gamma}{V}{A + B} \\
		\WellTypedTerm{\Gamma, z : A + B}{M}{\underline{C}}
	}{
        \begin{split}
            \TermEqual{\Gamma}{&\Case{V}{z}{\underline{C}}{x}{A}{M[\mathbf{inl}_{A+B}\ x/z]}{y}{B}{M[\mathbf{inr}_{A+B}\ y/z]}\\&}{M[V/z]}{\underline{C}[V/z]}
        \end{split}
    }
\end{mathpar}

\subsection{Semantics}
\subsubsection{Contexts}
\begin{mathpar}
	\inferrule{ }{\llbracket \emptyctx \rrbracket = 1}
	\and
	\inferrule{
		\llbracket \Gamma; A \rrbracket \in \category{E}_{\llbracket \Gamma \rrbracket} \\
		x \notin \mathrm{Vars}(\Gamma)
	}{\llbracket \Gamma, x : A \rrbracket = \ComprehensionFunctor{\llbracket \Gamma; A \rrbracket}}
\end{mathpar}

\subsubsection{Types}
\begin{mathpar}
    \inferrule{
        b : A \to \mathrm{Type} \\
        \llbracket \emptyctx; A \rrbracket \in \category{E}_1 \\
        \llbracket b \rrbracket \in \category{E}_{\ComprehensionFunctor{\llbracket \emptyctx; A \rrbracket}} \\
        \llbracket \Gamma; V \rrbracket : 1 \llbracket \Gamma \rrbracket \to \pullback{!_{\llbracket \Gamma \rrbracket}} \llbracket \emptyctx; A \rrbracket
    }{
        \llbracket \Gamma; b(V) \rrbracket = \pullback{(s \llbracket \Gamma; V \rrbracket)} \pullback{\ComprehensionFunctor{\CartesianOver{!_{\llbracket \Gamma \rrbracket}}{\llbracket \emptyctx; A \rrbracket}}} \llbracket b \rrbracket
    }
    \and
    \inferrule{
        \llbracket \Gamma \rrbracket \in \category{B}
    }{
        \llbracket \Gamma; 1 \rrbracket = 1 \llbracket \Gamma \rrbracket
    }
    \and
    \inferrule{
        \llbracket \Gamma; A \rrbracket \in \category{E}_{\llbracket \Gamma \rrbracket} \\
        \llbracket \Gamma, x : A; B \rrbracket \in \category{E}_{\llbracket \Gamma, x : A \rrbracket}
    }{
        \llbracket \Gamma; \Sigma x {:} A. B \rrbracket = \coprod_{\llbracket \Gamma; A \rrbracket} \llbracket \Gamma, x : A; B \rrbracket
    }
    \and
    \inferrule{
        \llbracket \Gamma; \underline{C} \rrbracket
    }{
        \llbracket \Gamma; U \underline{C} \rrbracket = U \llbracket \Gamma; \underline{C} \rrbracket
    }
    \and
    \inferrule{
        \llbracket \Gamma; A \rrbracket
    }{
        \llbracket \Gamma; F A \rrbracket = F \llbracket \Gamma; A \rrbracket
    }
    \and
    \inferrule{
        \llbracket \Gamma; A \rrbracket \\
        \llbracket \Gamma, x : A; \underline{C} \rrbracket
    }{
        \llbracket \Gamma; \CompProdType{x}{A}{\underline{C}} \rrbracket = \prod_{\llbracket \Gamma; A \rrbracket} \llbracket \Gamma, x : A; \underline{C} \rrbracket
    }
    \and
    \inferrule{
		\llbracket \Gamma; A \rrbracket \\
		\llbracket \Gamma; B \rrbracket
	}{\llbracket \Gamma; A + B \rrbracket = \llbracket \Gamma; A \rrbracket + \llbracket \Gamma; B \rrbracket}
\end{mathpar}

\subsubsection{Value Terms}

\begin{mathpar}
	\inferrule{
		\llbracket \Gamma, x : A \rrbracket
	}{
		\llbracket \Gamma, x : A; x \rrbracket \quad=\quad
		{\begin{tikzcd}[row sep=scriptsize]
			1 \llbracket \Gamma, x : A \rrbracket \ar[d, "\eta"] \\
			\pullback{\pi_{\llbracket \Gamma; A \rrbracket}} \coprod_{\llbracket \Gamma; A \rrbracket} 1 \llbracket \Gamma, x : A \rrbracket \ar[d, "\pullback{\pi_{\llbracket \Gamma; A \rrbracket}} \mathbf{fst}"] \\
			\pullback{\pi_{\llbracket \Gamma; A \rrbracket}} \llbracket \Gamma; A \rrbracket
		\end{tikzcd}}
	}
\end{mathpar}
\begin{mathpar}
	\inferrule{
		\llbracket \Gamma_1, x : A, \Gamma_2; B \rrbracket \\
		\llbracket \Gamma_1, x : A, \Gamma_2; x \rrbracket : 1 \llbracket \Gamma_1, x : A, \Gamma_2 \rrbracket \to A
	}{
		\llbracket \Gamma_1, x : A, \Gamma_2; y : B; x \rrbracket \quad=\quad
		{\begin{tikzcd}[row sep=scriptsize]
			1 \llbracket \Gamma_1, x : A, \Gamma_2, y : B \rrbracket \ar[d, equal] \\
			\pullback{\pi_{\llbracket \Gamma_1, x : A, \Gamma_2; B \rrbracket}} 1 \llbracket \Gamma_1, x : A, \Gamma_2 \rrbracket \ar[d, "{\pullback{\pi_{\llbracket \Gamma_1, x : A, \Gamma_2; B \rrbracket}} \llbracket \Gamma_1, x : A, \Gamma_2; x \rrbracket}"] \\
			\pullback{\pi_{\llbracket \Gamma_1, x : A, \Gamma_2; B \rrbracket}} A
		\end{tikzcd}}
	}
\end{mathpar}
\begin{mathpar}
	\inferrule{
		\llbracket \Gamma \rrbracket \in \category{B}
	}{
		\llbracket \Gamma; * \rrbracket =
		{\begin{tikzcd}[row sep=scriptsize]
			1 \llbracket \Gamma \rrbracket \ar[d, "\identity{1 \llbracket \Gamma \rrbracket}"] \\
			1 \llbracket \Gamma \rrbracket
		\end{tikzcd}}
	}
\end{mathpar}
\begin{mathpar}
	\inferrule{
		\llbracket \Gamma \rrbracket \\
		\llbracket c \rrbracket : 1 \to \llbracket \emptyctx; \mathrm{ty}(c) \rrbracket
	}{
		\llbracket \Gamma; c \rrbracket =
		{\begin{tikzcd}[row sep=scriptsize]
			1 \llbracket \Gamma \rrbracket \ar[d, "\pullback{!} \llbracket c \rrbracket"] \\
			\pullback{!} \llbracket \emptyctx; \mathrm{ty}(c) \rrbracket
		\end{tikzcd}}
	}
\end{mathpar}
\begin{mathpar}
	\inferrule{
		\llbracket \Gamma; V \rrbracket : 1 \llbracket \Gamma \rrbracket \to \llbracket \Gamma; A \rrbracket \\
		\llbracket \Gamma; W \rrbracket : 1 \llbracket \Gamma \rrbracket \to \pullback{(s \llbracket \Gamma; V \rrbracket)} \llbracket \Gamma, x : A; B \rrbracket
	}{
		\llbracket \Gamma; \DTuple{V}{W}{x}{A}{B} \rrbracket \quad=\quad
		{\begin{tikzcd}[row sep=scriptsize]
			1 \llbracket \Gamma \rrbracket \ar[d, "\llbracket \Gamma; W \rrbracket"] \\
			\pullback{(s \llbracket \Gamma; V \rrbracket)} \llbracket \Gamma, x : A; B \rrbracket \ar[d, "\pullback{(s \llbracket \Gamma; V \rrbracket)} \eta"] \\
			\pullback{(s \llbracket \Gamma; V \rrbracket)} \pullback{\pi_{\llbracket \Gamma; A \rrbracket}} \coprod_{\llbracket \Gamma; A \rrbracket} \llbracket \Gamma, x : A; B \rrbracket \ar[d, equal] \\
			\coprod_{\llbracket \Gamma; A \rrbracket} \llbracket \Gamma, x : A; B \rrbracket
		\end{tikzcd}}
	}
\end{mathpar}
\begin{mathpar}
	\inferrule{\llbracket \Gamma; M \rrbracket : 1 \llbracket \Gamma \rrbracket \to U \underline{C}}{
		\llbracket \Gamma; \thunk{M} \rrbracket = \llbracket \Gamma; M \rrbracket
	}
\end{mathpar}

\begin{mathpar}
	\inferrule{
		\llbracket \Gamma; V \rrbracket : 1 \llbracket \Gamma \rrbracket \to \llbracket \Gamma; A \rrbracket \\
		\llbracket \Gamma; B \rrbracket
	}{
		\llbracket \Gamma; \mathbf{inl}_{A + B}\ V \rrbracket \quad=\quad
		{\begin{tikzcd}[row sep=scriptsize]
			1 \llbracket \Gamma \rrbracket \ar[d, "{\llbracket \Gamma; V \rrbracket}"] \\
			\llbracket \Gamma; A \rrbracket \ar[d, "\iota_1"] \\
			\llbracket \Gamma; A \rrbracket + \llbracket \Gamma; B \rrbracket
		\end{tikzcd}}
	}
\end{mathpar}
\begin{mathpar}
	\inferrule{
		\llbracket \Gamma; V \rrbracket : 1 \llbracket \Gamma \rrbracket \to \llbracket \Gamma; B \rrbracket \\
		\llbracket \Gamma; A \rrbracket
	}{
		\llbracket \Gamma; \mathbf{inr}_{A + B}\ V \rrbracket \quad=\quad
		{\begin{tikzcd}[row sep=scriptsize]
			1 \llbracket \Gamma \rrbracket \ar[d, "{\llbracket \Gamma; V \rrbracket}"] \\
			\llbracket \Gamma; B \rrbracket \ar[d, "\iota_2"] \\
			\llbracket \Gamma; A \rrbracket + \llbracket \Gamma; B \rrbracket
		\end{tikzcd}}
	}
\end{mathpar}

\subsubsection{Computation Terms}
\begin{mathpar}
	\inferrule{
		\llbracket \Gamma; V \rrbracket : 1 \llbracket \Gamma \rrbracket \to A
	}{
		\llbracket \Gamma; \return{V} \rrbracket =
		{\begin{tikzcd}[row sep=scriptsize]
			1 \llbracket \Gamma \rrbracket \ar[d, "\llbracket \Gamma; V \rrbracket"] \\
			A \ar[d, "\eta_A"] \\
			U F A
		\end{tikzcd}}
	}
\end{mathpar}
\begin{mathpar}
	\inferrule{
		\llbracket \Gamma; M \rrbracket : 1 \llbracket \Gamma \rrbracket \to U F \llbracket \Gamma; A \rrbracket \\
		\llbracket \Gamma, x : A; N \rrbracket : 1 \llbracket \Gamma, x : A \rrbracket \to U \pullback{\pi_{\llbracket \Gamma; A \rrbracket}} \llbracket \Gamma; \underline{C} \rrbracket
	}{
		\llbracket \Gamma; \SeqComp{M}{x}{A}{\underline{C}}{N} \rrbracket \quad=\quad
		{\begin{tikzcd}[row sep=scriptsize]
			1 \llbracket \Gamma \rrbracket \ar[d, "\llbracket \Gamma; M \rrbracket"] \\
			U F \llbracket \Gamma; A \rrbracket \ar[d, "{U F \phi (\llbracket \Gamma, x : A; N \rrbracket)}"] \\
			U F U \llbracket \Gamma; \underline{C} \rrbracket \ar[d, "U \epsilon"] \\
			U \llbracket \Gamma; \underline{C} \rrbracket
		\end{tikzcd}}
	}
\end{mathpar}
\begin{mathpar}
	\inferrule{
		\llbracket \Gamma; V \rrbracket : 1 \llbracket \Gamma \rrbracket \to U \llbracket \Gamma; \underline{C} \rrbracket
	}{
		\llbracket \Gamma; \force{\underline{C}}{V} \rrbracket =
		{\begin{tikzcd}[row sep=scriptsize]
			1 \llbracket \Gamma \rrbracket \ar[d, "\llbracket \Gamma; V \rrbracket"] \\
			U \llbracket \Gamma; \underline{C} \rrbracket
		\end{tikzcd}}
	}
\end{mathpar}
\begin{mathpar}
	\inferrule{
		\llbracket \Gamma; V \rrbracket : 1 \llbracket \Gamma \rrbracket \to \coprod_{\llbracket \Gamma; A \rrbracket} \llbracket \Gamma, x : A; B \rrbracket \\
		\llbracket \Gamma, x : A, y : B; M \rrbracket : 1 \llbracket \Gamma, x : A, y : B \rrbracket \to U \pullback{\kappa} \llbracket \Gamma, z : \Sigma x {:} A. B; \underline{C} \rrbracket
	}{
		\llbracket \Gamma; \PatternMatch{V}{x}{A}{y}{B}{z}{\underline{C}}{M} \rrbracket \quad=\quad
		{\begin{tikzcd}[row sep=scriptsize]
			1 \llbracket \Gamma \rrbracket \ar[d, equal] \\
			\pullback{(s \llbracket \Gamma; V \rrbracket)} \pullback{(\kappa^{-1})} 1 \llbracket \Gamma, x : A, y : B \rrbracket \ar[d, "{\pullback{(s \llbracket \Gamma; V \rrbracket)} \pullback{(\kappa^{-1})} \llbracket \Gamma, x : A, y : B; M \rrbracket}"] \\
			\pullback{(s \llbracket \Gamma; V \rrbracket)} \pullback{(\kappa^{-1})} U \pullback{\kappa} \llbracket \Gamma, z : \Sigma x{:}A. B; \underline{C} \rrbracket \ar[d, equal] \\
			U \pullback{(s \llbracket \Gamma; V \rrbracket)} \llbracket \Gamma, z : \Sigma x{:}A. B; \underline{C} \rrbracket
		\end{tikzcd}}
	}
\end{mathpar}
\begin{mathpar}
	\inferrule{
		\llbracket \Gamma, x : A; M \rrbracket : 1 \llbracket \Gamma, x : A \rrbracket \to U \underline{C}
	}{
		\llbracket \Gamma; \LambdaAbs{x}{A}{M} \rrbracket \quad=\quad
		{\begin{tikzcd}[row sep=scriptsize]
			1 \llbracket \Gamma \rrbracket \ar[d, "\eta"] \\
			\prod_{\llbracket \Gamma; A \rrbracket} \pullback{\pi_{\llbracket \Gamma; A \rrbracket}} 1 \llbracket \Gamma \rrbracket \ar[d, equal] \\
			\prod_{\llbracket \Gamma; A \rrbracket} 1 \llbracket \Gamma, x : A \rrbracket \ar[d, "{\prod \llbracket \Gamma, x : A; M \rrbracket}"] \\
			\prod_{\llbracket \Gamma; A \rrbracket} U \underline{C} \ar[d, "\zeta^{-1}_{\llbracket \Gamma; A \rrbracket}"] \\
			U \prod_{\llbracket \Gamma; A \rrbracket} \underline{C}
		\end{tikzcd}}
	}
\end{mathpar}
\begin{mathpar}
	\inferrule{
		\llbracket \Gamma; M \rrbracket : 1 \llbracket \Gamma \rrbracket \to U \prod_{\llbracket \Gamma; A \rrbracket} \llbracket \Gamma, x : A; \underline{C} \rrbracket \\
		\llbracket \Gamma; V \rrbracket : 1 \llbracket \Gamma \rrbracket \to \llbracket \Gamma; A \rrbracket
	}{
		\llbracket \Gamma; \App{M}{V}{x}{A}{\underline{C}} \rrbracket \quad=\quad
		{\begin{tikzcd}[row sep=scriptsize]
			1 \llbracket \Gamma \rrbracket \ar[d, "\llbracket \Gamma; M \rrbracket"] \\
			U \prod_{\llbracket \Gamma; A \rrbracket} \llbracket \Gamma, x : A; \underline{C} \rrbracket \ar[d, equal] \\
			U \pullback{(s \llbracket \Gamma; V \rrbracket)} \pullback{\pi_{\llbracket \Gamma; A \rrbracket}} \prod_{\llbracket \Gamma; A \rrbracket} \llbracket \Gamma, x : A; \underline{C} \rrbracket \ar[d, "U \pullback{(s \llbracket \Gamma; V \rrbracket)} \epsilon"] \\
			U \pullback{(s \llbracket \Gamma; V \rrbracket)} \llbracket \Gamma, x : A; \underline{C} \rrbracket
		\end{tikzcd}}
	}
\end{mathpar}
\begin{mathpar}
	\inferrule{
		\llbracket \Gamma; V \rrbracket : 1 \llbracket \Gamma \rrbracket \to \llbracket \Gamma; A \rrbracket + \llbracket \Gamma; B \rrbracket \\
		\llbracket \Gamma, x : A; M \rrbracket : 1 \llbracket \Gamma, x : A \rrbracket \to U \pullback{\ComprehensionFunctor{\iota_1}} \llbracket \Gamma, z : A + B; \underline{C} \rrbracket \\
		\llbracket \Gamma, y : B; N \rrbracket : 1 \llbracket \Gamma, y : B \rrbracket \to U \pullback{\ComprehensionFunctor{\iota_2}} \llbracket \Gamma, z : A + B; \underline{C} \rrbracket
	}{
		\llbracket \Gamma; \Case{V}{z}{\underline{C}}{x}{A}{M}{y}{B}{N} \rrbracket \quad=\quad
		{\begin{tikzcd}[row sep=scriptsize]
			1 \llbracket \Gamma \rrbracket \ar[d, equal] \\
			\pullback{(s \llbracket \Gamma; V \rrbracket)} 1 \llbracket \Gamma, z : A + B \rrbracket \ar[d, "\pullback{(s \llbracket \Gamma; V \rrbracket)} {[\llbracket \Gamma, x : A; M \rrbracket, \llbracket \Gamma, y : B; N \rrbracket]}"] \\
			\pullback{(s \llbracket \Gamma; V \rrbracket)} U \llbracket \Gamma, z : A + B; \underline{C} \rrbracket \ar[d, equal] \\
			U \pullback{(s \llbracket \Gamma; V \rrbracket)} \llbracket \Gamma, z : A + B; \underline{C} \rrbracket
		\end{tikzcd}}
	}
\end{mathpar}

\section{Refinement Type System}
\subsection{Typing Rules}

\subsubsection{Well-Formed Contexts}
\begin{mathpar}
    \inferrule{ }{\WellFormedContext{\emptyctx}}
    \and
    \inferrule{
        \WellFormedContext{\Gamma} \\
        \WellFormedType{\Gamma}{A} \\
        x \notin \mathrm{Vars}(\Gamma)
    }{\WellFormedContext{\Gamma, x : A}}
\end{mathpar}

\subsubsection{Context Subtyping}
\begin{mathpar}
    \inferrule{ }{\ContextSubtyping{\emptyctx}{\emptyctx}}
    \and
    \inferrule{
        \ContextSubtyping{\Gamma_1}{\Gamma_2} \\
        \Subtyping{\Gamma_1}{A}{B} \\
        \WellFormedType{\Gamma_2}{B} \\
        x \notin \mathrm{Vars}(\Gamma_1) \cup \mathrm{Vars}(\Gamma_2)
    }{
        \ContextSubtyping{\Gamma_1, x : A}{\Gamma_2, x : B}
    }
\end{mathpar}

\subsubsection{Subtyping}
\paragraph{Reflexivity.}
\begin{mathpar}
    \inferrule{
        \WellFormedType{\Gamma}{A}
    }{
        \Subtyping{\Gamma}{A}{A}
    }
    \and
    \inferrule{
        \WellFormedType{\Gamma}{\underline{C}}
    }{
        \Subtyping{\Gamma}{\underline{C}}{\underline{C}}
    }
\end{mathpar}

\paragraph{Transitivity.}
\begin{mathpar}
    \inferrule{
        \Subtyping{\Gamma}{A_1}{A_2} \\
        \Subtyping{\Gamma}{A_2}{A_3}
    }{
        \Subtyping{\Gamma}{A_1}{A_3}
    }
    \and
    \inferrule{
        \Subtyping{\Gamma}{\underline{C}_1}{\underline{C}_2} \\
        \Subtyping{\Gamma}{\underline{C}_2}{\underline{C}_3}
    }{
        \Subtyping{\Gamma}{\underline{C}_1}{\underline{C}_3}
    }
\end{mathpar}

\subsubsection{Subsumption}
\begin{mathpar}
    \inferrule{
        \WellFormedType{\Gamma_2}{A} \\
        \ContextSubtyping{\Gamma_1}{\Gamma_2}
    }{
        \WellFormedType{\Gamma_1}{A}
    }
    \and
    \inferrule{
        \WellFormedType{\Gamma_2}{\underline{C}} \\
        \ContextSubtyping{\Gamma_1}{\Gamma_2}
    }{
        \WellFormedType{\Gamma_1}{\underline{C}}
    }
    \and
    \inferrule{
        \Subtyping{\Gamma_2}{A}{B} \\
        \ContextSubtyping{\Gamma_1}{\Gamma_2}
    }{
        \Subtyping{\Gamma_1}{A}{B}
    }
    \and
    \inferrule{
        \Subtyping{\Gamma_2}{\underline{C}}{\underline{D}} \\
        \ContextSubtyping{\Gamma_1}{\Gamma_2}
    }{
        \Subtyping{\Gamma_1}{\underline{C}}{\underline{D}}
    }
    \and
    \inferrule{
        \ContextSubtyping{\Gamma_1}{\Gamma_2} \\
        \WellTypedTerm{\Gamma_2}{V}{A} \\
        \Subtyping{\Gamma_1}{A}{B}
    }{
        \WellTypedTerm{\Gamma_1}{V}{B}
    }
    \and
    \inferrule{
        \ContextSubtyping{\Gamma_1}{\Gamma_2} \\
        \WellTypedTerm{\Gamma_2}{M}{\underline{C}} \\
        \Subtyping{\Gamma_1}{\underline{C}}{\underline{D}}
    }{
        \WellTypedTerm{\Gamma_1}{M}{\underline{D}}
    }
\end{mathpar}

\subsubsection{Variables}
\begin{mathpar}
    \inferrule{
        \WellFormedContext{\Gamma_1, x : A, \Gamma_2}
    }{
        \WellTypedTerm{\Gamma_1, x : A, \Gamma_2}{x}{A}
    }
    \and
    \inferrule{
        \WellFormedContext{\Gamma_1, x : \{ v : b(V) \mid p \}, \Gamma_2}
    }{
        \WellTypedTerm{\Gamma_1, x : \{ v : b(V) \mid p \}, \Gamma_2}{x}{\{ v : b(V) \mid v = x \}}
    }
\end{mathpar}

\subsubsection{Value constants}
\begin{mathpar}
    \inferrule{
        \WellFormedContext{\Gamma} \\
        \WellFormedType{\emptyctx}{\mathrm{ty}(c)}
    }{
        \WellTypedTerm{\Gamma}{c_{\ElimRefinement{\mathrm{ty}(c)}}}{\mathrm{ty}(c)}
    }
\end{mathpar}

\subsubsection{Unit Type}
\begin{mathpar}
    \inferrule{\WellFormedContext{\Gamma}}{\WellTypedTerm{\Gamma}{*}{\{ v : 1 \mid \top \}}}
    \and
    \inferrule{
        \WellFormedContext{\Gamma} \\
        \ElimRefinement{\Gamma}, v : 1 \vdash p : \mathrm{Prop}
    }{\WellFormedType{\Gamma}{\{ v : 1 \mid p \}}}
    \and
    \inferrule{
        \WellFormedContext{\Gamma} \\
        \Gamma; v : 1 \mid p \vdash q
    }{\Subtyping{\Gamma}{\{ v : 1 \mid p \}}{\{ v : 1 \mid q \}}}
\end{mathpar}

\subsubsection{Refinement Types}
\begin{mathpar}
    \inferrule{
        \WellFormedContext{\Gamma} \\
        b : \underlying{A} \to \mathrm{Type} \\
        \WellFormedType{\ElimRefinement{\Gamma}}{b(V)} \\
        \ElimRefinement{\Gamma}, v : b(V) \vdash p : \mathrm{Prop}
    }{
        \WellFormedType{\Gamma}{\{ v : b(V) \mid p \}}
    }
    \and
    \inferrule{
        \WellFormedContext{\Gamma} \\
        \TypeEqual{\ElimRefinement{\Gamma}}{b(V)}{b(W)} \\
        \Gamma; v : b(V) \mid p \vdash q
    }{
        \Gamma \vdash \{ v : b(V) \mid p \} <: \{ v : b(W) \mid q \}
    }
\end{mathpar}

\subsubsection{Value $\Sigma$-Types}
\begin{mathpar}
    \inferrule{
        \WellFormedType{\Gamma}{A} \\
        \WellFormedType{\Gamma, x : A}{B}
    }{\WellFormedType{\Gamma}{\ValueSumType{x}{A}{B}}}
    \and
    \inferrule{
        \WellTypedTerm{\Gamma}{V}{A} \\
        \WellFormedType{\Gamma, x : A}{B} \\
        \WellTypedTerm{\Gamma}{W}{B[V/x]}
    }{\WellTypedTerm{\Gamma}{\langle V, W \rangle}{\ValueSumType{x}{A}{B}}}
    \and
    \inferrule{
        \WellTypedTerm{\Gamma}{V}{\ValueSumType{x}{A}{B}} \\
        \WellFormedType{\Gamma, z : \ValueSumType{x}{A}{B}}{\underline{C}} \\
        \WellTypedTerm{\Gamma, x : A, y : B}{M}{\underline{C}[\langle x, y \rangle / z]}
    }{\WellTypedTerm{\Gamma}{\PatternMatch{V}{x}{\ElimRefinement{A}}{y}{\ElimRefinement{B}}{z}{\ElimRefinement{\underline{C}}}{M}}{\underline{C}[V/z]}}
    \and
    \inferrule{
        \Subtyping{\Gamma}{A_1}{A_2} \\
        \WellFormedType{\Gamma, x : A_2}{B_2} \\
        \Subtyping{\Gamma, x : A_1}{B_1}{B_2}
    }{
        \Subtyping{\Gamma}{\Sigma x : A_1. B_1}{\Sigma x : A_2. B_2}
    }
\end{mathpar}

\subsubsection{Thunked Computation}
\begin{mathpar}
    \inferrule{
        \WellFormedType{\Gamma}{\underline{C}}
    }{
        \WellFormedType{\Gamma}{U\underline{C}}
    }
    \and
    \inferrule{
        \WellTypedTerm{\Gamma}{M}{\underline{C}}
    }{
        \WellTypedTerm{\Gamma}{\mathbf{thunk}\ M}{U \underline{C}}
    }
    \and
    \inferrule{
        \WellTypedTerm{\Gamma}{V}{U \underline{C}}
    }{
        \WellTypedTerm{\Gamma}{\mathbf{force}_{\ElimRefinement{\underline{C}}}\ V}{\underline{C}}
    }
    \and
    \inferrule{
        \Subtyping{\Gamma}{\underline{C}_1}{\underline{C}_2}
    }{
        \Subtyping{\Gamma}{U \underline{C}_1}{U \underline{C}_2}
    }
\end{mathpar}

\subsubsection{Return}
\begin{mathpar}
    \inferrule{
        \WellFormedType{\Gamma}{A}
    }{
        \WellFormedType{\Gamma}{F A}
    }
    \and
    \inferrule{
        \WellTypedTerm{\Gamma}{V}{A}
    }{
        \WellTypedTerm{\Gamma}{\return{V}}{F A}
    }
    \and
    \inferrule{
        \WellTypedTerm{\Gamma}{M}{F A} \\
        \WellFormedType{\Gamma}{\underline{C}} \\
        \WellTypedTerm{\Gamma, x : A}{N}{\underline{C}}
    }{
        \WellTypedTerm{\Gamma}{\SeqComp{M}{x}{\ElimRefinement{A}}{\ElimRefinement{\underline{C}}}{N}}{\underline{C}}
    }
    \and
    \inferrule{
        \Subtyping{\Gamma}{A_1}{A_2}
    }{
        \Subtyping{\Gamma}{F A_1}{F A_2}
    }
\end{mathpar}

\subsubsection{Computational $\Pi$-Types}
\begin{mathpar}
    \inferrule{
        \WellFormedType{\Gamma}{A} \\
        \WellFormedType{\Gamma, x : A}{\underline{C}}
    }{
        \WellFormedType{\Gamma}{\CompProdType{x}{A}{\underline{C}}}
    }
    \and
    \inferrule{
        \Subtyping{\Gamma}{A_2}{A_1} \\
        \WellFormedType{\Gamma, x : A_1}{\underline{C}_1} \\
        \Subtyping{\Gamma, x : A_2}{\underline{C}_1}{\underline{C}_2}
    }{
        \Subtyping{\Gamma}{\CompProdType{x}{A_1}{\underline{C}_1}}{\CompProdType{x}{A_2}{\underline{C}_2}}
    }
    \and
    \inferrule{
        \WellTypedTerm{\Gamma, x : A}{M}{\underline{C}}
    }{
        \WellTypedTerm{\Gamma}{\lambda x : \ElimRefinement{A}. M}{\CompProdType{x}{A}{\underline{C}}}
    }
    \and
    \inferrule{
        \WellFormedType{\Gamma, x : A}{\underline{C}} \\
        \WellTypedTerm{\Gamma}{M}{\CompProdType{x}{A}{\underline{C}}} \\
        \WellTypedTerm{\Gamma}{V}{A}
    }{
        \WellTypedTerm{\Gamma}{M(V)_{(x : \ElimRefinement{A}). \ElimRefinement{\underline{C}}}}{\underline{C}[V/x]}
    }
\end{mathpar}

\subsubsection{Fibred Coproduct Types}
\begin{mathpar}
    \inferrule{
        \WellFormedType{\Gamma}{A} \\
        \WellFormedType{\Gamma}{B}
    }{
        \WellFormedType{\Gamma}{A + B}
    }
    \and
    \inferrule{
        \WellTypedTerm{\Gamma}{V}{A} \\
        \WellFormedType{\Gamma}{B}
    }{
        \WellTypedTerm{\Gamma}{\mathbf{inl}_{A + B}\ V}{A + B}
    }
    \and
    \inferrule{
        \WellTypedTerm{\Gamma}{V}{B} \\
        \WellFormedType{\Gamma}{A}
    }{
        \WellTypedTerm{\Gamma}{\mathbf{inr}_{A + B}\ V}{A + B}
    }
    \and
    \inferrule{
		\WellFormedType{\Gamma, z : A + B}{\underline{C}} \\
		\WellTypedTerm{\Gamma}{V}{A + B} \\
		\WellTypedTerm{\Gamma, x : A}{M}{\underline{C}[\mathbf{inl}_{A+B}\ x/z]} \\
		\WellTypedTerm{\Gamma, y : B}{N}{\underline{C}[\mathbf{inr}_{A+B}\ y/z]}
    }{\WellTypedTerm{\Gamma}{\mathbf{case}\ V\ \mathbf{of}_{z.\underline{C}}\ (\mathbf{inl}\ (x : A) \mapsto M, \mathbf{inr}\ (y : B) \mapsto N)}{\underline{C}[V/z]}}
    \and
    \inferrule{
        \Subtyping{\Gamma}{A_1}{A_2} \\
        \Subtyping{\Gamma}{B_1}{B_2}
    }{
        \Subtyping{\Gamma}{A_1 + B_1}{A_2 + B_2}
    }
\end{mathpar}

\subsection{Semantics}

\subsubsection{Contexts}
\begin{mathpar}
	\inferrule{ }{\llbracket \emptyctx \rrbracket = 1 \in \category{P}_1}
	\and
	\inferrule{
		\llbracket \Gamma; A \rrbracket \in \RefinedCCompC{\category{E}}{\category{P}}_{\llbracket \Gamma \rrbracket} \\
		x \notin \mathrm{Vars}(\Gamma)
	}{\llbracket \Gamma, x : A \rrbracket = \ComprehensionFunctor{\llbracket \Gamma; A \rrbracket}}
\end{mathpar}

\subsubsection{Types}
\begin{mathpar}
    \inferrule{
        \llbracket \Gamma \rrbracket \in \category{P}_I \\
        \llbracket \ElimRefinement{\Gamma}; b(V) \rrbracket \in \category{E}_I \\
        \llbracket \ElimRefinement{\Gamma}, v : b(V) \vdash p \rrbracket \in \category{P}_{\ComprehensionFunctor{X}}
    }{
        \llbracket \Gamma; \{ v : b(V) \mid p \} \rrbracket = \big(\llbracket \ElimRefinement{\Gamma}; b(V) \rrbracket, \llbracket \Gamma \rrbracket, \pullback{\pi_{\llbracket \ElimRefinement{\Gamma}; b(V) \rrbracket}} \llbracket \Gamma \rrbracket \land \llbracket \ElimRefinement{\Gamma}, v : b(V) \vdash p \rrbracket\big)
    }
    \and
    \inferrule{
		\llbracket \Gamma \rrbracket \in \category{P} \\
		\llbracket \ElimRefinement{\Gamma}, v : 1 \vdash p \rrbracket \in \category{P}_{\ComprehensionFunctor{1 q \llbracket \Gamma \rrbracket}}
    }{\llbracket \Gamma; \{ v : 1 \mid p \} \rrbracket = \big(1 q \llbracket \Gamma \rrbracket, \llbracket \Gamma \rrbracket, \pullback{\pi_{1 q \llbracket \Gamma \rrbracket}} \llbracket \Gamma \rrbracket \land \llbracket \ElimRefinement{\Gamma}, v : 1 \vdash p \rrbracket\big)}
    \and
    \inferrule{
        \llbracket \Gamma; A \rrbracket \\
        \llbracket \Gamma, x : A; B \rrbracket
    }{
        \llbracket \Gamma; \Sigma x {:} A. B \rrbracket = \coprod_{\llbracket \Gamma; A \rrbracket} \llbracket \Gamma, x : A; B \rrbracket
    }
    \and
    \inferrule{
        \llbracket \Gamma; \underline{C} \rrbracket
    }{
        \llbracket \Gamma; U \underline{C} \rrbracket = \dot{U} \llbracket \Gamma; \underline{C} \rrbracket
    }
    \and
    \inferrule{
        \llbracket \Gamma; A \rrbracket
    }{
        \llbracket \Gamma; F A \rrbracket = \dot{F} \llbracket \Gamma; A \rrbracket
    }
    \and
    \inferrule{
        \llbracket \Gamma; A \rrbracket \\
        \llbracket \Gamma, x : A; \underline{C} \rrbracket
    }{
        \llbracket \Gamma; \CompProdType{x}{A}{\underline{C}} \rrbracket = \prod_{\llbracket \Gamma; A \rrbracket} \llbracket \Gamma, x : A; \underline{C} \rrbracket
    }
\end{mathpar}

\end{document}